\documentclass[aps,pra,twocolumn,showpacs,superscriptaddress]{revtex4-1}  
\usepackage{graphicx}  
\usepackage{dcolumn}   
\usepackage{bm}        
\usepackage{amssymb}   
\usepackage{amsmath}   
\usepackage{subfigure} 
\usepackage{hyperref} 
\hypersetup{
    colorlinks=true,
    linkcolor=blue,
    citecolor=blue,      
    urlcolor=blue,
    pdftitle={High-pressure phases of group II difluorides: polymorphism and superionicity},
    bookmarks=true,
}

\begin{document}

\title{High-pressure phases of group II difluorides: polymorphism and superionicity} 

\author{Joseph R.~Nelson} \email{jn336@cam.ac.uk}

\author{Richard J.~Needs} \affiliation{Theory of Condensed Matter Group,
  Cavendish Laboratory, J.~J.~Thomson Avenue, Cambridge CB3 0HE,
  United Kingdom}

\author{Chris J.~Pickard} \affiliation{Department of Materials Science
  and Metallurgy, University of Cambridge, 27 Charles Babbage Road,
  Cambridge CB3 0FS, United Kingdom} \affiliation{Advanced Institute
  for Materials Research, Tohoku University, 2-1-1 Katahira, Aoba,
  Sendai, 980-8577, Japan} \vskip 0.25cm

\date{\today}

\begin{abstract}
  We investigate the high-pressure behaviour of beryllium, magnesium
  and calcium difluorides using \textit{ab initio} random structure
  searching and density functional theory (DFT) calculations, over the
  pressure range 0$-$70 GPa. Beryllium fluoride exhibits extensive
  polymorphism at low pressures, and we find two new phases for this
  compound $-$ the silica moganite and CaCl$_2$ structures $-$ which
  are stable over the wide pressure range 12$-$57 GPa. For magnesium
  fluoride, our searching results show that the orthorhombic `O-I'
  TiO$_2$ structure ($Pbca$, $Z=8$) is stable for this compound
  between 40 and 44 GPa. Our searches find no new phases at the
  static-lattice level for calcium difluoride between 0 and 70 GPa;
  however, a phase with $P\overline{6}2m$ symmetry is close to
  stability over this pressure range, and our calculations predict
  that this phase is stabilised at high temperature. The
  $P\overline{6}2m$ structure exhibits an unstable phonon mode at
  large volumes which may signal a transition to a superionic state at
  high temperatures. The Group-II difluorides are isoelectronic to a
  number of other AB$_2$-type compounds such as SiO$_2$ and TiO$_2$,
  and we discuss our results in light of these similarities.
\end{abstract}


\maketitle

\section{\label{intro}Introduction}
The Group-II difluorides form materials with a wide variety of
technological uses.  BeF$_2$, and mixtures of it with further
fluorides and difluorides, are used to create glasses for use in
infrared photonics, which have excellent transmittance in the
UV. BeF$_2$ glass itself has a large bandgap of 13.8
eV \cite{Dumbaugh-1980,Williams-1981}. BeF$_2$ is chemically stable and
is employed as a mixture component in nuclear reactor molten salts,
where it is useful as a coolant, and is also capable of dissolving
fissile materials \cite{vanderMeer-2007}. MgF$_2$ is birefringent with
a wide wavelength transmission range, and is used in the manufacture
of optical components such as polarizers \cite{Appel-2002}. CaF$_2$
also offers a high transmittance across a wide range of wavelengths
and is used in optical systems, as well as an internal pressure
standard \cite{Daimon-2002,Hazen-1981}. MgF$_2$ and CaF$_2$ occur
naturally as the minerals sellaite and fluorite, therefore high
pressure modifications of these compounds are of interest in
geophysics.

Group-II difluorides have 16 valence electrons per formula unit, and
are isoelectronic to other AB$_2$ compounds of industrial or
geophysical significance, such as TiO$_2$ and SiO$_2$. As such, these
compounds share many similar crystal structures, albeit stable at
different pressures. Because of this structural similarity, Group-II
difluorides have been investigated as structural analogues of silica
phases \cite{Grocholski-2013}. BeF$_2$ is particularly similar to
silica at low pressures \cite{Dachille-1959}, as in addition to being
isoelectronic, the fluoride (F$^{-}$) and oxide (O$^{2-}$) ions have
similar radii and polarisabilities, and the Be/F and Si/O atomic radii
ratios are similar at $\approx$ 0.3 \cite{Walsh-2009}. MgF$_2$ has also
been explored as a model for higher pressure silica
phases \cite{Haines-2001}.

We are interested in the structures and phases of Be-, Mg- and CaF$_2$
at ambient and elevated pressures, and the implications of such phases
for other AB$_2$ compounds. Our approach to determining stable phases
in these compounds uses computational structure searching alongside
density functional theory (DFT) calculations, and we elect to explore
the pressure range 0$-$70 GPa.

\section{\label{methods}Methods}
The \textit{ab initio} random structure searching (AIRSS)
technique \cite{PN2011} is used to search for Group-II difluoride
structures at three pressures: 15, 30 and 60 GPa. AIRSS is a
stochastic method which generates structures randomly with a given
number of formula units. A minimum atom-atom separation is specified
for the generated structures, e.g., we set the Be-Be, Be-F, and F-F
minimum separations for the case of BeF$_2$. These separations are
chosen based on short AIRSS runs. We can also impose symmetry
constraints on our generated structures such that low symmetry
structures are not considered. This strategy tends to speed up the
searches because such low-symmetry structures are unlikely to have low
energies according to Pauling's principle 
\cite{Pauling_1929,Needs_Pickard_2016}, although we allocate part of
our searching time to check low symmetry structures, for
completeness. AIRSS has a proven track record of predicting structures
in a diverse variety of systems that have subsequently been verified
by experiment, such as in compressed silane, aluminium hydride,
high-pressure hydrogen sulfide and xenon
oxides \cite{PN2006,PN2007,Li-2016,Xenon_oxides_2016}.

We limit our searches to a maximum of 8 formula units (24 atoms) per
cell. In addition to AIRSS, we supplement our searches with a set of
15 known AB$_2$-type structures taken from a variety of compounds; see
the Appendix for a full list of these structures. Structures generated
by AIRSS or taken from known AB$_2$ compounds are relaxed to a
enthalpy minimum using a variable-cell geometry optimization
calculation. For this, we use density-functional theory (DFT) as
implemented in the \textsc{castep} plane-wave pseudopotential code
\cite{CS2005}, with internally-generated ultrasoft pseudopotentials
\cite{V1990} and the Perdew-Burke-Ernzerhof (PBE) exchange-correlation
functional \cite{PBE1996}. Our DFT calculations use a 800 eV
plane-wave basis set cutoff, and a Brillouin zone sampling density of
at least $2\pi\times0.04$ \AA$^{-1}$. Our calculations of phonon
frequencies use the quasiharmonic approximation (QHA)
\cite{KW2003,Carrier} and a finite-displacement supercell method, as
implemented in \textsc{castep}.

As well as searches for new BeF$_2$, MgF$_2$ and CaF$_2$ structures,
we have also carried out variable stoichiometry structure searching at
60 GPa, to examine the possibility of other thermodynamically stable
compositions. These searches predict that BeF$_2$ and MgF$_2$ are the
only stable stoichiometries at 60 GPa, while both CaF$_2$ and CaF$_3$
are stable at that pressure. Our predicted CaF$_3$ structure is cubic
with $Pm\overline{3}n$ symmetry, and the same structure has in fact
been previously predicted for high-pressure aluminium hydride
(AlH$_3$) \cite{AlH3}. We have not examined the properties of
high-pressure CaF$_3$ further, and in what follows we focus only on
the AB$_2$ difluoride stoichiometry. Convex hulls for our variable
stoichiometry searches are given in the Appendix.

Semilocal density functionals such as PBE typically underestimate the
calculated band gap in materials. In order to capture the optical
properties of the group II difluorides, we perform optical bandgap
calculations with the non-local Heyd-Scuseria-Ernzerhof (HSE06)
functional \cite{Heyd-2003}. This functional incorporates 25\% screened
exchange and is expected to generally improve the accuracy of bandgap
calculations carried out with DFT, though at a greater computational
cost. We adopt the following calculation strategy: structures are
first relaxed with the PBE functional, norm-conserving
pseudopotentials \cite{footnote_1}, a basis set cutoff of 1600 eV and a
relatively sparse Brillouin zone sampling of $2\pi\times0.1$
\AA$^{-1}$. The HSE06 functional is then used with PBE-relaxed
geometries for a self-consistent calculation of the electronic bands
of the structure. Our stress calculations indicate that the use of the
HSE06 functional with PBE geometries gives rise to small forces of
$\lesssim$ 0.1 eV/\AA\ on each atom. Electronic density of states
calculations are then performed using the \textsc{OptaDOS}
code \cite{optados1,optados2,optados3}. Additional information on the
electronic structure calculations in this study can be found in the
Appendix.

\section{Beryllium difluoride}
\begin{figure*}[htp]
  \centering
  \subfigure{\includegraphics[scale=0.326,clip]{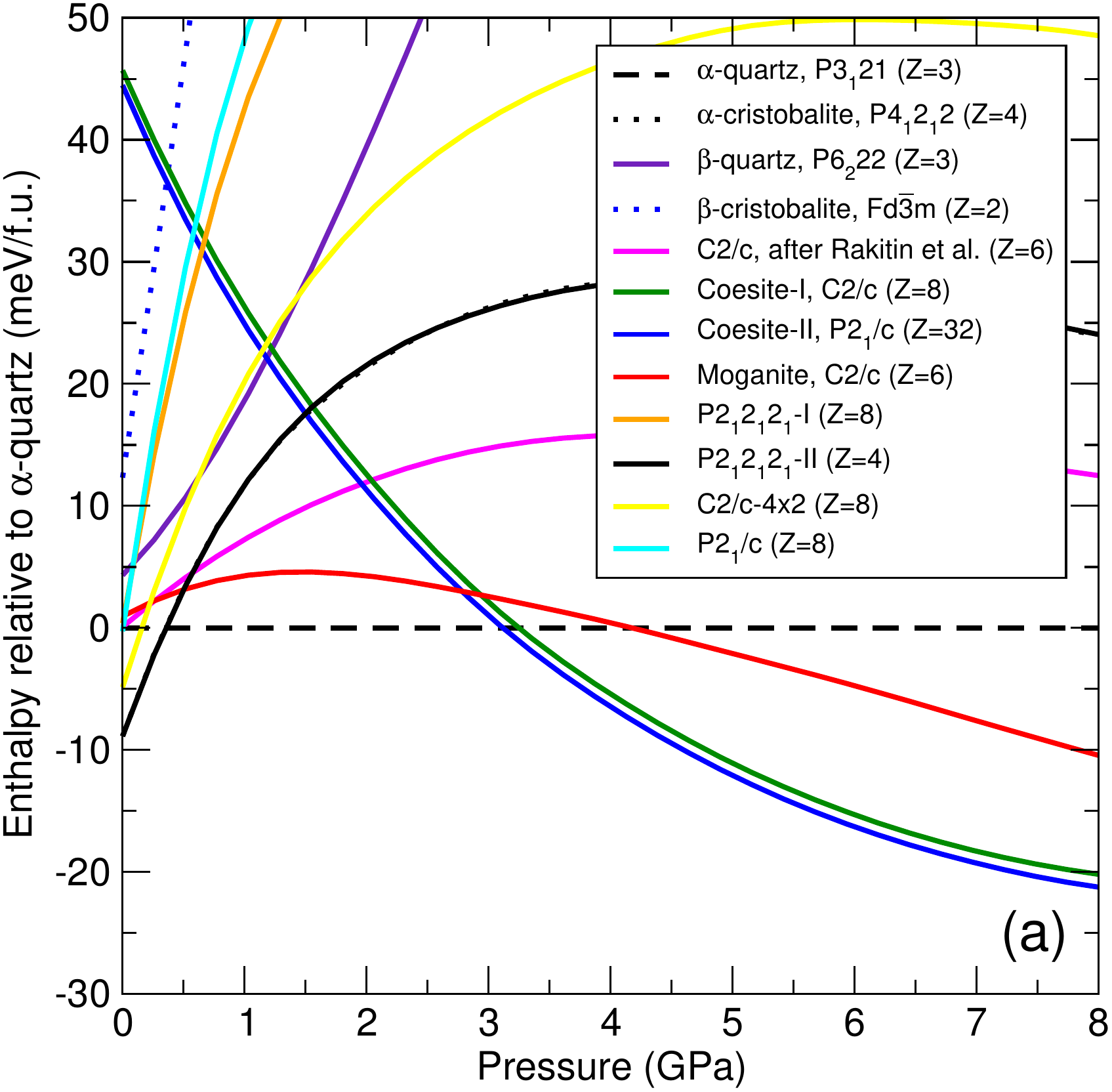}}\quad
  \subfigure{\includegraphics[scale=0.326,clip]{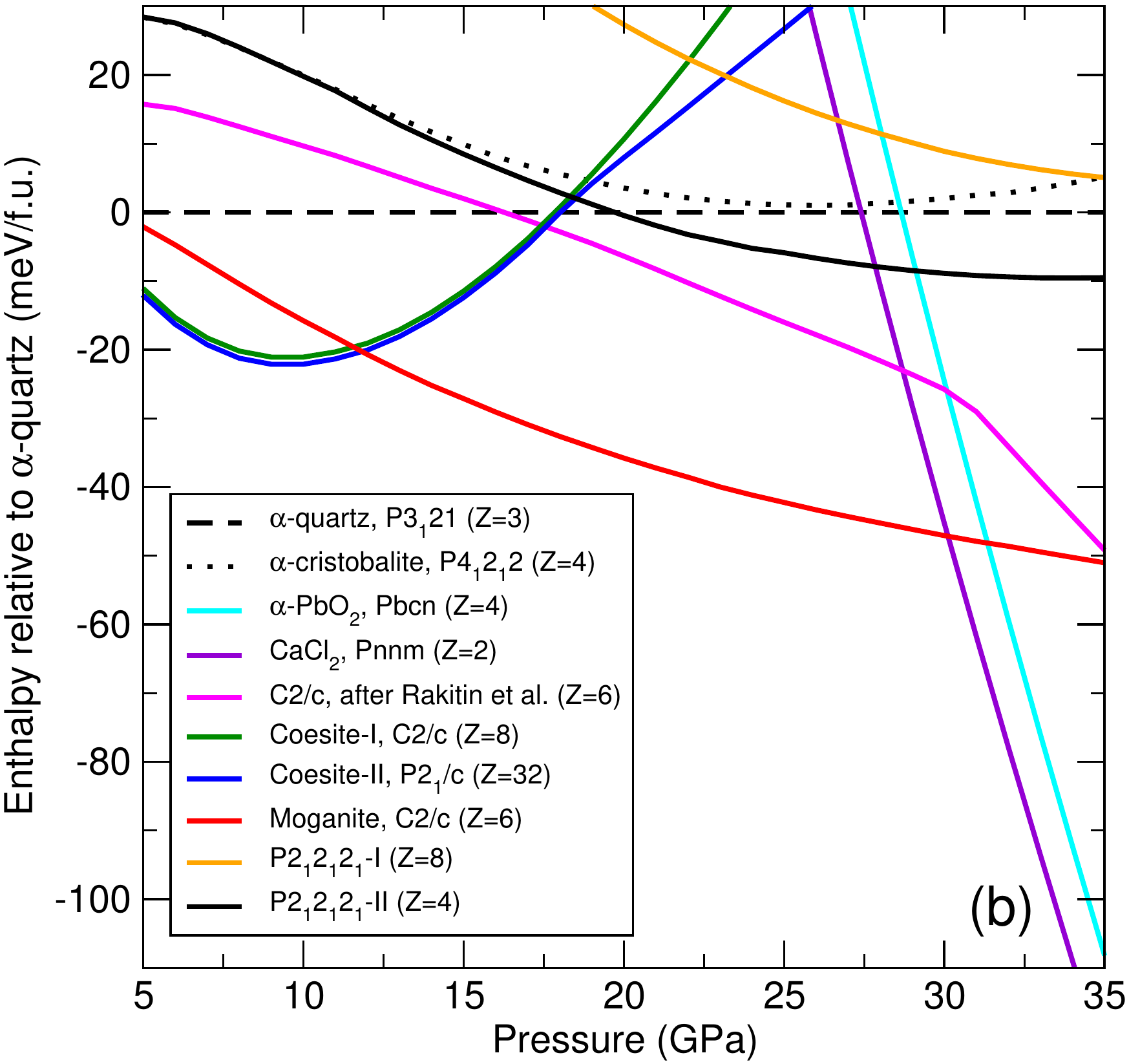}}\quad
  \subfigure{\includegraphics[scale=0.326,clip]{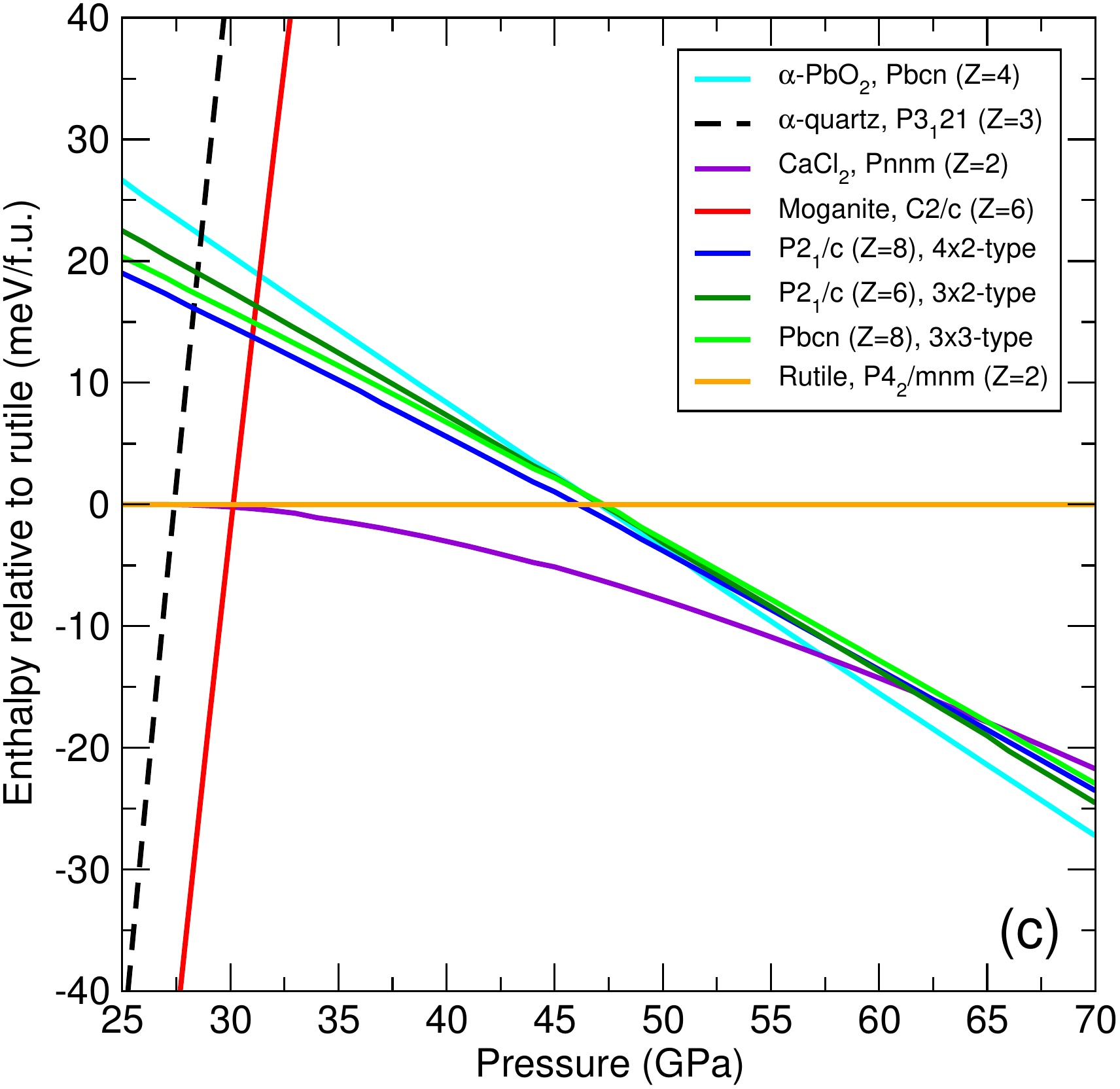}}
  \caption{\label{fig:BeF2_Enthalpies} Static lattice enthalpies for
    BeF$_2$ calculated with the PBE functional over the pressure
    ranges (a) 0$-$8 GPa, (b) 5$-$35 GPa and (c) 25$-$70 GPa. In (a) and (b),
    enthalpies are shown relative to the $\alpha$-quartz phase, while
    in (c) they are shown relative to the rutile phase.}
\end{figure*}

\begin{figure}
\centering
  \includegraphics[scale=0.15]{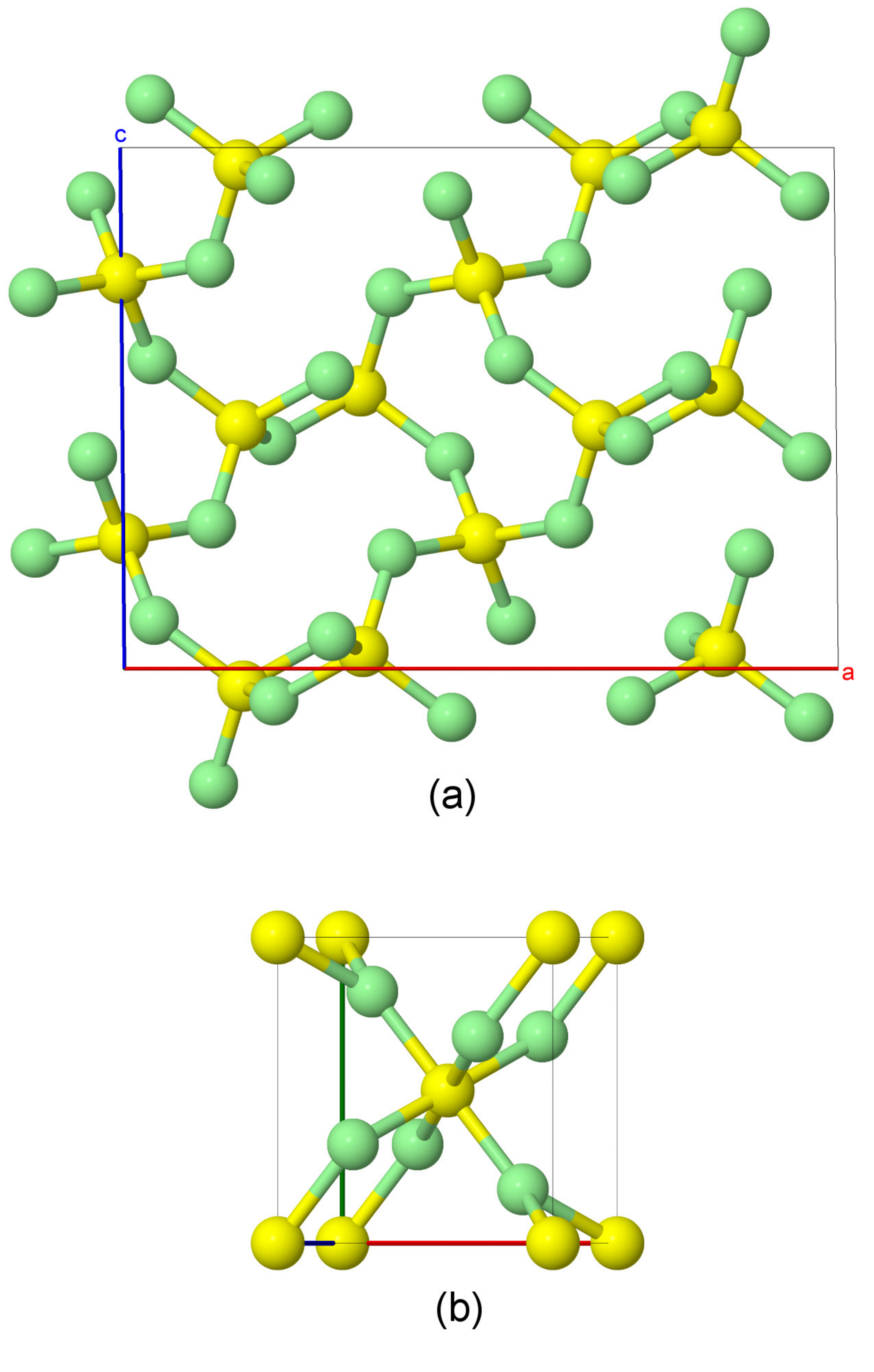}
  \caption{\label{fig:moganite_20GPa} (a) BeF$_2$ in the $C2/c$
    moganite phase at \mbox{20 GPa} with four-fold coordinated Be
    atoms, viewed down the $b$ (and in this case, monoclinic)
    axis. The lattice $a$ and $c$ axes point horizontal and almost
    vertical in the page, respectively. (b) BeF$_2$ in the $Pnnm$
    CaCl$_2$ phase at 50 GPa, with six-fold coordinated Be
    atoms. Beryllium atoms in yellow, fluorine atoms in green. Lattice
    parameters and atomic positions for these structures are given in
    the Appendix.}
\end{figure}

\subsection{Low$-$pressure results}
Beryllium fluoride has several thermodynamically accessible phases
which have been obtained in experimental studies. At temperatures
below its melting point (820 K) and pressures at or below atmospheric
pressure, the $\alpha$-quartz, $\beta$-quartz, $\alpha$-cristobalite,
$\beta$-cristobalite and glass phases of BeF$_2$ have been prepared
under various conditions \cite{Wright-1988,Ghalsasi-2010}. 

Our results from structure searching calculations at low pressures
(0$-$8 GPa) are summarised in Fig.~\ref{fig:BeF2_Enthalpies}(a), which
shows the static-lattice enthalpies of several BeF$_2$ structures as a
function of pressure. Beryllium fluoride shows extensive polymorphism
at low pressures; we find at least 10 structures for BeF$_2$ (all
those shown in Fig.~\ref{fig:BeF2_Enthalpies}(a), except Coesite-I and Coesite-II)
lying within about 20 meV/f.u. of one another at 0 GPa. The $\alpha$-
and $\beta$-quartz, and $\alpha$- and $\beta$-cristobalite structures
are part of this set of 10 structures. Within our current
calculational framework (DFT$-$PBE), the $\alpha$-cristobalite
structure is lowest in enthalpy at 0 GPa, lying 8.9 meV/f.u. below
$\alpha$-quartz, but other low enthalpy structures are also of
interest as they can be stabilised by temperature. There is some
experimental evidence for as-yet-unknown high-temperature low-pressure
BeF$_2$ phases \cite{Jackson-1976}, and metastable structures produced
in our searches provide useful reference crystal structures that could
be matched to available experimental data at elevated temperatures. We
remark here that the PBE functional may not provide a completely
accurate energy ordering for these structures at low pressure. For
example in silica, DFT$-$PBE also predicts that the
$\alpha$-cristobalite structure has a lower enthalpy than the
$\alpha$-quartz structure at 0 GPa. However, high pressure phase
transition pressures in silica calculated with PBE are in excellent
agreement with accurate Quantum Monte$-$Carlo
calculations \cite{Driver-2010}.

A recent study by Rakitin \textit{et al.} found a $C2/c$ structure for
BeF$_2$ which is close in enthalpy to $\alpha$-quartz at 0
GPa \cite{Rakitin-2015}. Results using AIRSS produce five further
polymorphs for BeF$_2$, labelled `Moganite', $P2_12_12_1$-I,
$P2_12_12_1$-II, $C2/c-$4$\times$2 and $P2_1/c$ in
Fig.~\ref{fig:BeF2_Enthalpies}(a), which have static-lattice
enthalpies of +0.9, +0.4, $-$8.9, $-$4.9 and $-$0.3 meV/f.u. relative
to the $\alpha$-quartz phase at 0 GPa. We discuss each of these in
more detail below. 

\textbf{Moganite}. The `Moganite' polymorph, which turned up in our
searches (and was also included in our set of 15 AB$_2$ structures $-$
see the Appendix), is BeF$_2$ in the silica moganite structure with
Si$\rightarrow$Be and O$\rightarrow$F. This has space group $C2/c$ and
6 formula units of BeF$_2$ in the primitive cell.

\textbf{P2$_1$2$_1$2$_1$-II}. At the pressures shown in
Fig.~\ref{fig:BeF2_Enthalpies}(a), the $P2_12_12_1$-II polymorph has
space group symmetry $P4_32_12$ (\#96), and is actually an enantiomer
of $\alpha$-cristobalite. A mirror-image transformation
$(x,y,z)\rightarrow (-y,-x,z)$, i.e. a reflection of atomic positions
in the (110) plane, relates the two structures. This situation is
entirely analogous to SiO$_2$, where rigid-unit phonon modes (RUMs)
distort the $\beta$-cristobalite structure to one of either $P4_12_12$
symmetry (our $\alpha$-cristobalite) or to $P4_32_12$ symmetry (our
$P2_12_12_1$-II). Coh \textit{et al.} \cite{Coh-2008} employ the
notation $\tilde{\alpha}_1$ and $\tilde{\alpha}_1'$ respectively for
these two cristobalite structures. Enthalpy-pressure curves for the
$\alpha$-cristobalite and $P2_12_12_1$-II structures lie on top of one
another in Fig.~\ref{fig:BeF2_Enthalpies}(a), and our calculations
give the overall change in enthalpy going from $\beta$-cristobalite to
$\tilde{\alpha}_1$ or $\tilde{\alpha}_1'$ as and $-21.2$ meV/BeF$_2$;
the equivalent quantity is $-34.4$ meV/SiO$_2$ in silica. Above 11
GPa, the $P2_12_12_1$-II structure distorts to the lower space group
symmetry $P2_12_12_1$ (\#19), and the enthalpy curves for it and the
$\alpha$-cristobalite structure start to diverge
(Fig.~\ref{fig:BeF2_Enthalpies}(b)).

\textbf{Open framework structures}. The \mbox{$P2_12_12_1$-I},
$C2/c-$4$\times$2 and $P2_1/c$ structures in
Fig.~\ref{fig:BeF2_Enthalpies}(a) are low-density polymorphs which are
previously unreported in BeF$_2$. The notation `$C2/c-$4$\times$2'
refers to the fact that this structure is a low-pressure variant of
the `$P2_1/c$ ($Z$=8), 4$\times$2-type' structure shown in
Fig.~\ref{fig:BeF2_Enthalpies}(c). Our DFT$-$PBE calculations show
that these polymorphs are energetically relevant in silica, where they
lie $5.5$, $15.7$ and $5.9$ meV/SiO$_2$ in energy below
$\alpha$-quartz at 0 GPa, which suggests them as likely silica
polymorphs as well. The monoclinic angle in the $P2_1/c$ structure is
close to 90$^{\circ}$, hence we also investigate a higher symmetry
version of this structure with $Pnma$ symmetry, whose enthalpy is
found to be $2.2$ meV/SiO$_2$ below $\alpha$-quartz (this structure is
not shown in Fig.~\ref{fig:BeF2_Enthalpies}(a)). Calculated lattice
parameters and atomic positions for these structures are provided in
the Appendix.

We query the ICSD \cite{ICSD} and International Zeolite Association
(IZA) \cite{IZA} databases to check if these four structures are
already known in SiO$_2$. Our $C2/c-$4$\times$2 structure matches
\#75654 in the ICSD, which is `Structure 8' in the simulated annealing
structure prediction work of Boisen \textit{et al.}
\cite{Boisen-1994}, after the latter is relaxed using DFT. However, we
find no matching SiO$_2$ structures in these databases for our
$P2_12_12_1$-I, $P2_1/c$ and $Pnma$ structures.  Analysis of these
three structures using the TOPOS code \cite{TOPOS} shows that they are
all of the same topological type as the ABW zeolite. They have
framework densities of 18.9, 18.6 and 18.5 Si/1000\AA$^3$
(cf. $\alpha$-quartz, with 24.9 Si/1000\AA$^3$), while the ABW silica
structure itself is slightly less dense at 17.6 Si/1000\AA$^3$, with
an enthalpy 3.9 meV/SiO$_2$ above $\alpha$-quartz. While BeF$_2$ and
SiO$_2$ share many chemical similarities (as mentioned in
Sec.~\ref{intro}), the Be$-$F bond is much weaker than the Si$-$O
bond, resulting in a lower melting point and hardness for BeF$_2$
\cite{Wright-1988}. Nevertheless, our results show that BeF$_2$ also
supports open framework zeolite-like structures, and highlights the
utility of searching for potential zeolite structures in model systems
such as BeF$_2$. The large number of polymorphs we encounter lying
close to one another in energy suggest BeF$_2$ as a potential
tetrahedral framework material.

\subsection{\label{BeF2_highp}High$-$pressure results}
Our structure searching calculations show that the application of
pressure (0.4 GPa) favours the $\alpha$-quartz phase, as seen in
Fig.~\ref{fig:BeF2_Enthalpies}(a). Between 3.1 and 3.3 GPa, the silica
`coesite-I' or `coesite-II' structure \cite{Cernok-2014} with $Z$=8 or
32 then becomes the lowest-enthalpy structure for BeF$_2$. We find
that over the pressure range 0-18 GPa, the coesite-I and II structures
are nearly identical. The coesite-II structure is close to a supercell
of coesite-I, but the atomic positions in coesite-II deviate slightly
from those expected for a perfect supercell, and the coesite-II
structure lies consistently about 1 meV/BeF$_2$ below coesite-I over
this pressure range. The coesite-I phase has been found in
experimental studies on BeF$_2$, at 3 GPa and $\approx$1100 K
\cite{Jackson-1976}. Previous work \cite{Rakitin-2015} reported that a
new structure of $C2/c$ symmetry then becomes stable between 18 and 27
GPa; our calculations instead show that BeF$_2$ is most stable in the
silica moganite structure, which also has $C2/c$ symmetry, between
11.6 and 30.1 GPa (see Fig.~\ref{fig:BeF2_Enthalpies}(b)). Above 30.1
GPa, we find that the orthorhombic CaCl$_2$ structure with space group
$Pnnm$ becomes stable (Fig.~\ref{fig:BeF2_Enthalpies}(b)), eventually
giving way to the denser $\alpha$-PbO$_2$ structure above 57.5 GPa
(Fig.~\ref{fig:BeF2_Enthalpies}(c)). The moganite and CaCl$_2$
structures are depicted in Fig.~\ref{fig:moganite_20GPa}.

The enthalpy-pressure curve for the CaCl$_2$ structure emerges
smoothly from that for the rutile structure (space group $P4_2/mnm$),
which is also the case in silica, where a ferroelastic phase
transition occurs between these two structures near 50
GPa \cite{Togo-2008}. For BeF$_2$ at the static lattice level, our
calculations exclude the stability of the rutile structure over the
pressure range 0$-$70 GPa, though we note that this phase lies only a
fraction of a meV per BeF$_2$ higher in enthalpy than the CaCl$_2$ phase
at 30.1 GPa, as seen in Fig.~\ref{fig:BeF2_Enthalpies}(b). Earlier
studies \cite{Yu-2013, Rakitin-2015} have already examined a number of
the structures discussed here; however, the stability of BeF$_2$ in
the moganite and CaCl$_2$ structures is new, and according to our
calculations dominates the high-pressure phases of BeF$_2$ over the
pressure range 11.6$-$57.5 GPa.

Fig.~\ref{fig:BeF2_Enthalpies}(c) shows a band of three
enthalpy-pressure curves, labelled \mbox{$Pbcn$ ($Z$=8)},
\mbox{$P2_1/c$ ($Z$=6)} and \mbox{$P2_1/c$ ($Z$=8)}, whose energy lies
in close proximity to the $\alpha$-PbO$_2$ curve. These three
structures emerged from our searches and have a similar, but slightly
lower density than the $\alpha$-PbO$_2$ phase for BeF$_2$. They are
close to stability at 60 GPa, but are not predicted to be stable over
the pressure range 0$-$70 GPa. We identify these phases as members of
the class of silica polymorphs introduced by Teter \textit{et
  al}. \cite{Teter-1998}, which are a set of structures described as
intermediaries to the CaCl$_2$ and $\alpha$-PbO$_2$ silica phases. Our
\mbox{$Pbcn$ ($Z$=8)} and \mbox{$P2_1/c$ ($Z$=6)} structures
correspond to the `$3\times 3$' and `$3\times 2$' structure types,
while our \mbox{$P2_1/c$ ($Z$=8)} structure is not explicitly
discussed in Ref.~\cite{Teter-1998} and would be referred to as
`$4\times 2$' type. We will encounter these phases again in our
results for MgF$_2$.

\textbf{Summary}. To briefly summarise our search results for BeF$_2$, we predict the
following series of pressure-induced phase transitions at the
static-lattice level:
\begin{eqnarray*}
 &\alpha\mbox{-cristobalite}\:(P4_12_12)\: \xrightarrow{\mbox{0.4 GPa}}\: 
  \alpha\mbox{-quartz}\:(P3_121) \\
 &\xrightarrow{\mbox{3.1/3.3 GPa}}\: \mbox{Coesite-I/II}\:(C2/c)\: \\
 &\xrightarrow{\mbox{11.6 GPa}} \mbox{Moganite}\:(C2/c)
  \xrightarrow{\mbox{30.1 GPa}} \mbox{CaCl}_2\:(Pnnm)\: \\
 & \xrightarrow{\mbox{57.5 GPa}} \alpha\mbox{-PbO}_2\:(Pbcn),
\end{eqnarray*}
with the labelled arrows showing the calculated transition
pressures. Our searches also demonstrate numerous metastable
polymorphs for BeF$_2$.

\subsection{Optical bandgaps in BeF$_2$}
\begin{figure}[h]
  \includegraphics[scale=0.46,clip]{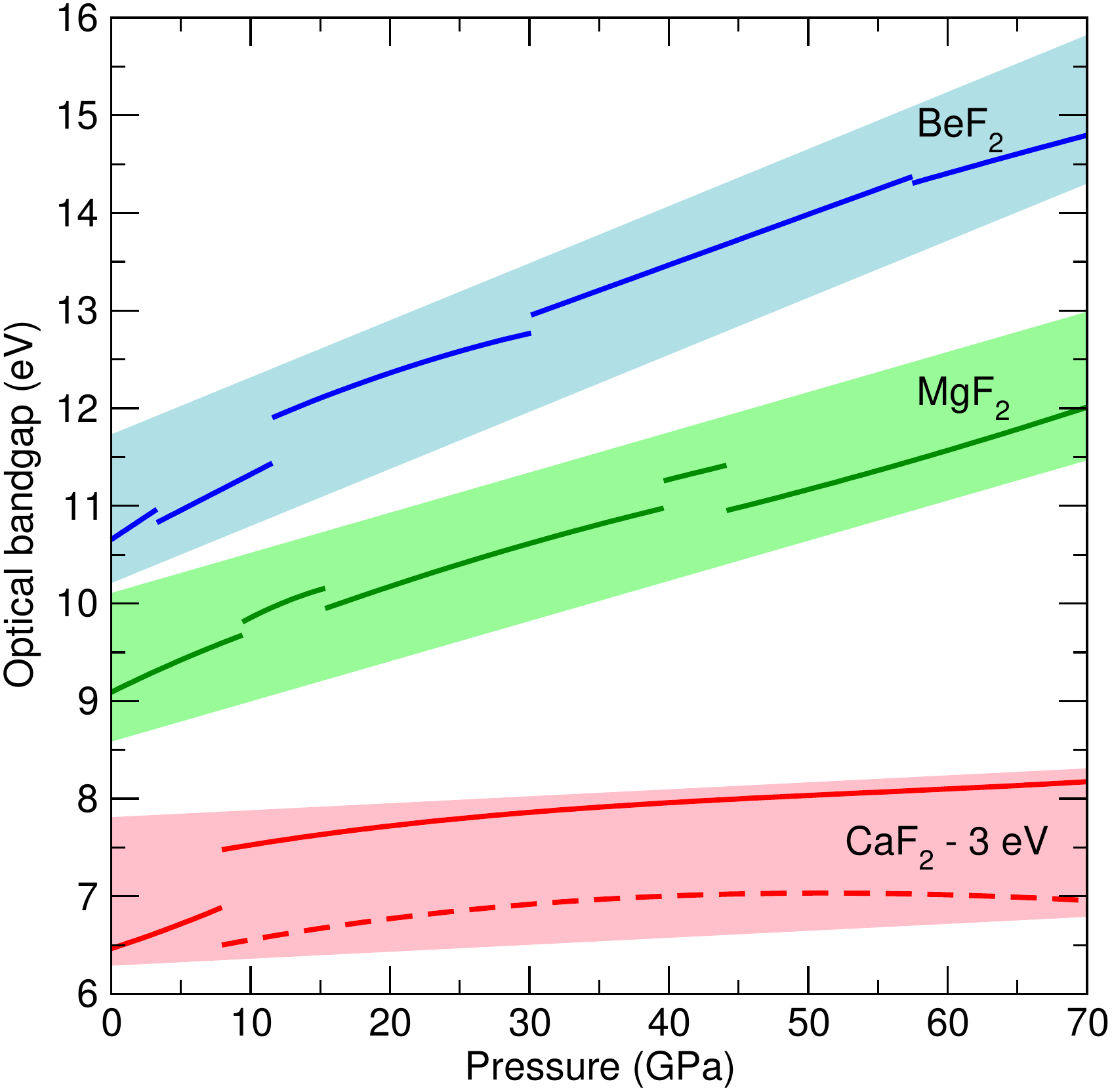}
  \caption{\label{fig:optical_bandgaps} Optical bandgaps in Be-, Mg-
    and CaF$_2$ as calculated using the HSE06 functional, over the
    pressure range 0$-$70 GPa. For visibility, the bandgaps in CaF$_2$
    have been shifted down by 3 eV. Discontinuities in the solid curves
    are due to phase transitions between different structures, while
    the shaded regions serve to guide the eye. The dashed curve in
    CaF$_2$ corresponds to the $P\overline{6}2m$ phase, which is not
    stable at the static lattice level but which we predict is
    stabilised by temperature.}
\end{figure}

As mentioned in Sec.~\ref{intro}, BeF$_2$ has a large bandgap at
ambient pressure and is used in a number of optical applications. We
examine the optical bandgap in BeF$_2$ as a function of pressure, with
the results shown in Fig.~\ref{fig:optical_bandgaps}. The optical gap
is found to be tunable, increasing by around 0.06 eV/GPa over the
pressure range 0$-$70 GPa. We expect BeF$_2$ to therefore maintain its
high UV transmittance with increasing pressure, with potentially
useful high pressure applications. The electronic density-of-states
(DOS) of the moganite and CaCl$_2$ phases are also given in the
Appendix.

\section{Magnesium difluoride}

\begin{figure}
\centering
  \includegraphics[scale=0.46,clip]{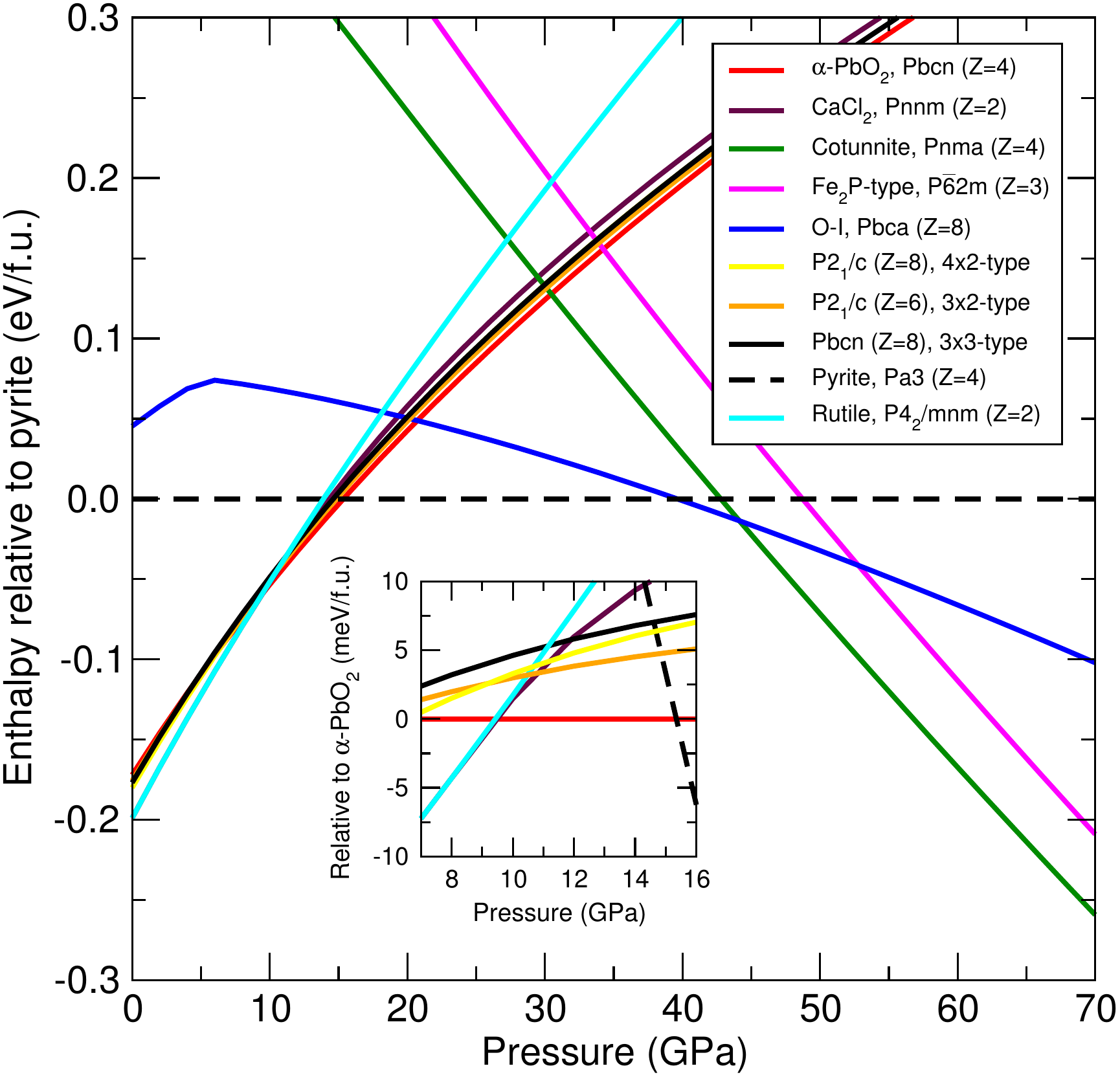}
  \caption{\label{fig:MgF_searchresults} Static-lattice enthalpies and
    results from structure searches on MgF$_2$ over the pressure range
    0$-$70 GPa. The inset plot shows the small enthalpy differences
    between a few phases in the vicinity of 10 GPa. Enthalpies are
    shown relative to the pyrite (main figure) and $\alpha$-PbO$_2$
    (inset) phases.}
\end{figure}

\begin{figure}
\centering
  \includegraphics[scale=0.15,clip]{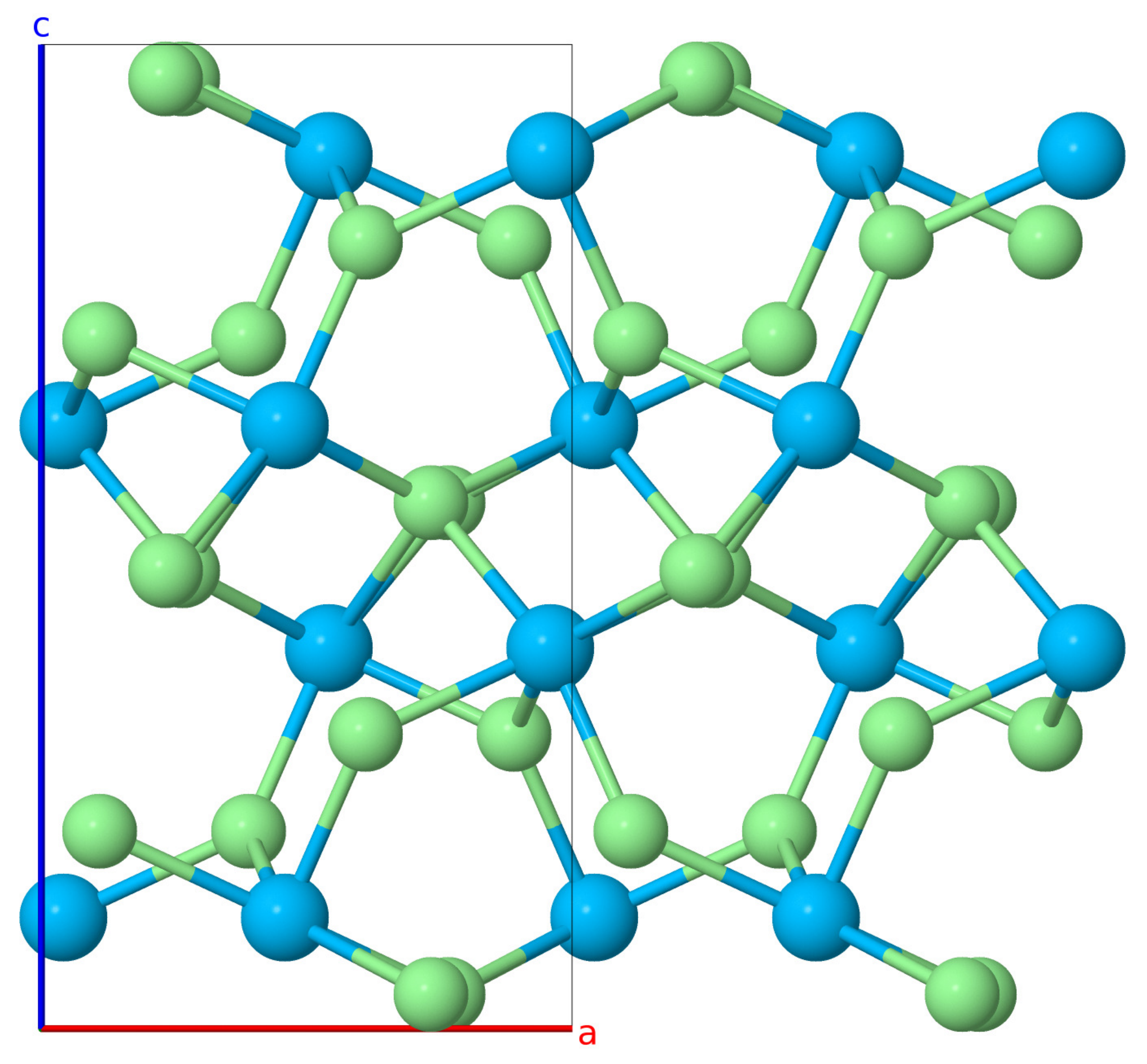}
  \caption{\label{fig:MgF2_Pbca} 2$\times$1$\times$1 slab of MgF$_2$
    in the $Pbca$ `orthorhombic-I' structure ($Z$=8), which we predict
    to be stable between 39.6 and 44.1 GPa. This view looks down the
    $b$ axis. Magnesium atoms in blue, fluorine atoms in green.}
\end{figure}

MgF$_2$ adopts the rutile $P4_2/mnm$ structure at room temperature and
pressure. X-ray diffraction experiments indicate a transformation to
the CaCl$_2$ structure at 9.1 GPa, then to a pyrite structure with
space group $Pa3$ and $Z$=4 near 14 GPa, and have also
recovered a mixture of $\alpha$-PbO$_2$ and rutile MgF$_2$ upon
decompression \cite{Haines-2001}. DFT calculations, including those of
the present study, actually show that the CaCl$_2$ structure is never
stable for MgF$_2$ and instead predict the $\alpha$-PbO$_2$ structure
to have a window of stability between 10 and 15 GPa, with the CaCl$_2$
structure slightly higher in enthalpy.

The results from our structure searches are given in
Fig.~\ref{fig:MgF_searchresults}. Based on our static-lattice results,
we predict the following sequence of stable structures and phase
transitions with rising pressure:
\begin{eqnarray*}
 &\mbox{Rutile}\:(P4_2/mnm)\:\xrightarrow{\mbox{9.4 GPa}} \:\alpha\mbox{-PbO}_2\:(Pbcn)\: \\
 & \xrightarrow{\mbox{15.4 GPa}} \mbox{Pyrite}\:(Pa3)\: \xrightarrow{\mbox{39.6 GPa}} \mbox{O-I}\:(Pbca)\: \\
& \xrightarrow{\mbox{44.1 GPa}}\: \mbox{Cotunnite}\:(Pnma).
\end{eqnarray*}

Previous theoretical studies \cite{Ozturk-2014,Kanchana-2003} have
already considered the rutile, $\alpha$-PbO$_2$, pyrite and cotunnite
phases of MgF$_2$. In the present work, we find that the $Pbca$ O-I
`orthorhombic-I' structure, which has been reported experimentally for
TiO$_2$ near 30 GPa \cite{Dubrovinskaia-2001}, is also stable for
MgF$_2$ between 39.6 and 44.1 GPa. This structure is depicted in
Fig.~\ref{fig:MgF2_Pbca}. Experimental studies on MgF$_2$ have indeed
reported an unidentified `Phase X' stable in the pressure range
49$-$53 GPa and at 1500$-$2500 K between the pyrite and cotunnite
phases \cite{Grocholski-2010}. Our enthalpy calculations identify the
O-I structure as the thermodynamically most likely candidate for Phase
X, though the authors of Ref.~\cite{Grocholski-2010} note some
difficulty in indexing x-ray diffraction data on Phase X to an
orthorhombic structure, possibly due to a mixture of phases being
present. We do not find any other energetically competitive structures
near 50 GPa.

Silica and its stable polymorphs are of paramount importance in
geophysics and planetary sciences. As well as a mineral in its own
right, it is expected to be formed from the breakdown of
post-perovskite MgSiO$_3$ at terapascal pressures. SiO$_2$ follows a
very similar set of phase transitions to MgF$_2$ with increasing
pressure \cite{Wu-2011}, with the Si coordination number rising from 6
in rutile at ambient pressures to a predicted 10 in an $I4/mmm$
structure near 10 TPa \cite{Lyle-2015}. As pointed out by previous
authors, several features of high pressure silica can readily be
modelled in MgF$_2$, but at much lower
pressures \cite{Haines-2001,Grocholski-2010}. For example, the
$\alpha$-PbO$_2$$\,\rightarrow\,$pyrite transition in SiO$_2$, which our
calculations find occurs at 217 GPa, takes place at a much lower
pressure of 15.4 GPa in MgF$_2$. Near 690 GPa and for $T\gtrsim 1000$
K, a pyrite$\,\rightarrow\,$cotunnite transition is also predicted for
SiO$_2$ \cite{Tsuchiya-2011}; the analogous transition occurs at 44.1
GPa in MgF$_2$.

As mentioned in Sec.~\ref{BeF2_highp}, Teter \textit{et
  al}. \cite{Teter-1998} have introduced a class of SiO$_2$
polymorphs intermediate to CaCl$_2$ and $\alpha$-PbO$_2$.
At least one member of this class of polymorphs has been synthesised
in SiO$_2$, the `$3\times2$' type $P2_1/c$ structure \cite{Haines-2001-II}. 
The \mbox{$P2_1/c$ ($Z$=6)}, \mbox{$P2_1/c$ ($Z$=8)} and \mbox{$Pbcn$ ($Z$=8)} 
structures of BeF$_2$ depicted in Fig.~\ref{fig:BeF2_Enthalpies}(c) 
are members of this class, and also turn up in our MgF$_2$
searches (Fig.~\ref{fig:MgF_searchresults} and its inset). Our
calculations show that these polymorphs are closest to stability near
10 GPa in MgF$_2$, compared to $\approx$100 GPa in SiO$_2$
suggesting that, as with other features of silica, they could be
studied experimentally at much lower pressures in MgF$_2$.

Fig.~\ref{fig:optical_bandgaps} shows the optical bandgap in MgF$_2$
as a function of pressure. Like BeF$_2$, the optical gap is tunable
with pressure, rising by about 0.04 eV/GPa over 0$-$70 GPa. We also
provide the electronic DOS of the O-I structure in the Appendix.

\section{Calcium difluoride}

\begin{figure}
\centering
  \includegraphics[scale=0.45,clip]{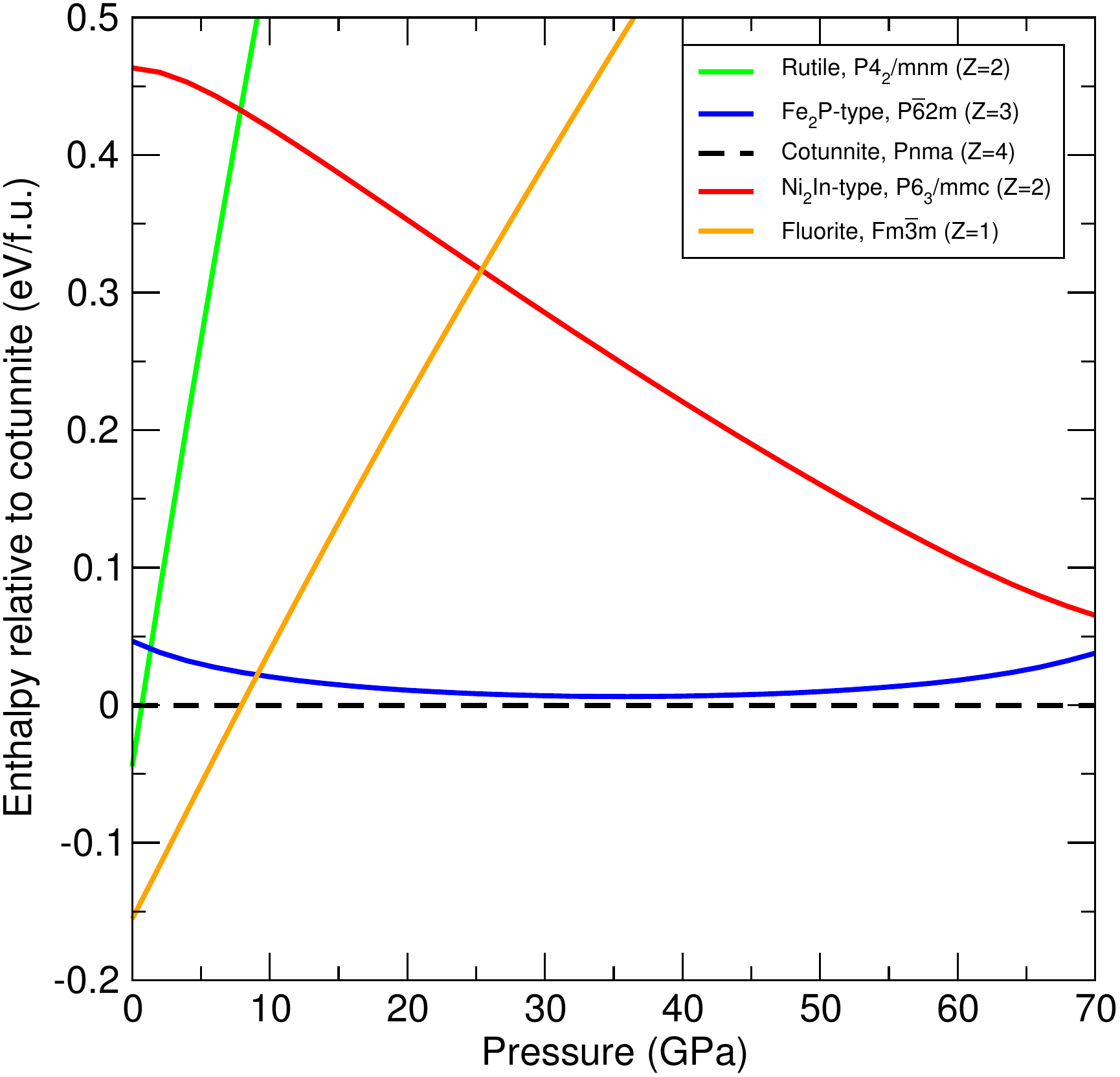}
  \caption{\label{fig:CaF_searchresults} Results from structure
    searches on CaF$_2$ over the pressure range 0$-$70 GPa:
    static-lattice enthalpies relative to the $Pnma$ ($\gamma$)
    CaF$_2$ phase.}
\end{figure}

\begin{figure}[h]
  \includegraphics[scale=0.11,clip]{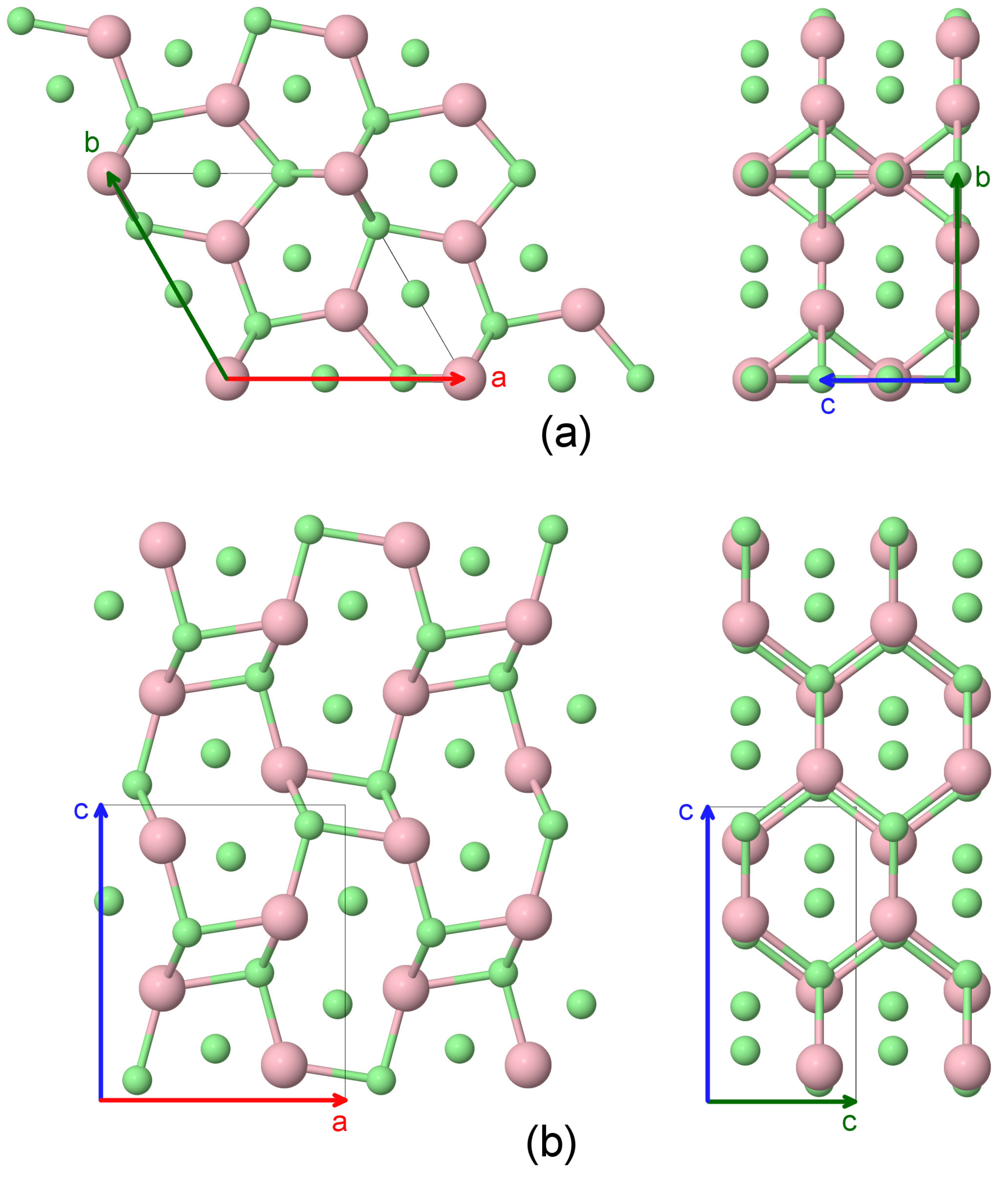}
  \caption{\label{fig:P62m_and_Pnma} 2$\times$2$\times$2 slabs of (a)
    our predicted $P6\overline{2}m$ structure, and (b) the $Pnma$
    structure (phase $\gamma$) of CaF$_2$. Calcium atoms are in red,
    fluorine atoms in green. In both (a) and (b), the right-hand view
    is obtained from the left-hand view by rotating the structure by
    90$^{\circ}$ about an axis running vertically up the page.}
\end{figure}

CaF$_2$ crystallizes in the cubic $Fm\overline{3}m$ `fluorite' structure
($Z$=$4$, $\alpha$-CaF$_2$) under ambient conditions. The compound has
a high-temperature phase above about 1400 K, known as $\beta$-CaF$_2$,
and melts near 1700 K at low pressures \cite{Mirwald-1977}. A
high-pressure modification above 8$-$10 GPa ($\gamma$-CaF$_2$) is also
known, with CaF$_2$ taking on the orthorhombic $Pnma$ cotunnite
structure ($Z$=$4$) \cite{Gerward-1992}.

The $\beta$ phase has attracted considerable interest because it
exhibits superionicity, with F$^{-}$ ions as the diffusing
species \cite{Gillan-1985}. A number of other compounds in the same
fluorite (or `anti-fluorite') variants of this structure, such as
Li$_2$O, are also superionic conductors \cite{Mulliner-2015}. Such
materials are of great technological interest, with applications in
solid-state battery design. A recent study has shown that the
superionic transition temperature in CaF$_2$ can be decreased through
applied stress \cite{Cazorla-2016}.

\subsection{Results from structure searching}
Our results from structure searching in CaF$_2$ are shown in
Fig.~\ref{fig:CaF_searchresults}. Unlike BeF$_2$ and MgF$_2$, we find
that the potential energy surface for CaF$_2$ is relatively simple,
with very few polymorphs for this compound over the pressure range
0$-$70 GPa. At the static lattice level of theory, we identify only the
sequence of stable phases and transitions:
\begin{eqnarray*}
\mbox{Fluorite }(Fm\overline{3}m) \xrightarrow{\mbox{7.9 GPa}} \mbox{Cotunnite }(Pnma).
\end{eqnarray*}

The calculated fluorite$\,\rightarrow\,$cotunnite transition pressure here
is in agreement with experimental results \cite{Gerward-1992}.
Experimental studies have also shown a transition from
$\gamma$-CaF$_2$ to an Ni$_2$In-type structure in the pressure range
63$-$79 GPa with laser heating \cite{Dorfman-2010}, consistent with the
convergence of the red and black-dashed curves in
Fig.~\ref{fig:CaF_searchresults}.

Our results in Fig.~\ref{fig:CaF_searchresults} reveal a hexagonal
phase for CaF$_2$ with $P\overline{6}2m$ symmetry which is close to
stability, lying only 6 meV/CaF$_2$ higher in enthalpy than the
$\gamma$ phase near 36 GPa. The enthalpy curves for the
$P\overline{6}2m$ and $\gamma$ phases in
Fig.~\ref{fig:CaF_searchresults} indicate that these two structures
have very similar densities, with $P\overline{6}2m$ slightly denser at
pressures below 36 GPa and becoming less dense than $\gamma$-CaF$_2$
at higher pressures. The $P\overline{6}2m$ phase has the Fe$_2$P
structure, which has also been predicted for SiO$_2$ at very high
pressures ($>$0.69 TPa) and low temperatures \cite{Tsuchiya-2011}. We
show the $\gamma$ and $P\overline{6}2m$ structures in
Fig.~\ref{fig:P62m_and_Pnma}.

\subsection{Pressure-temperature phase diagram for CaF$_2$}

\begin{figure}
\centering
  \includegraphics[scale=0.46,clip]{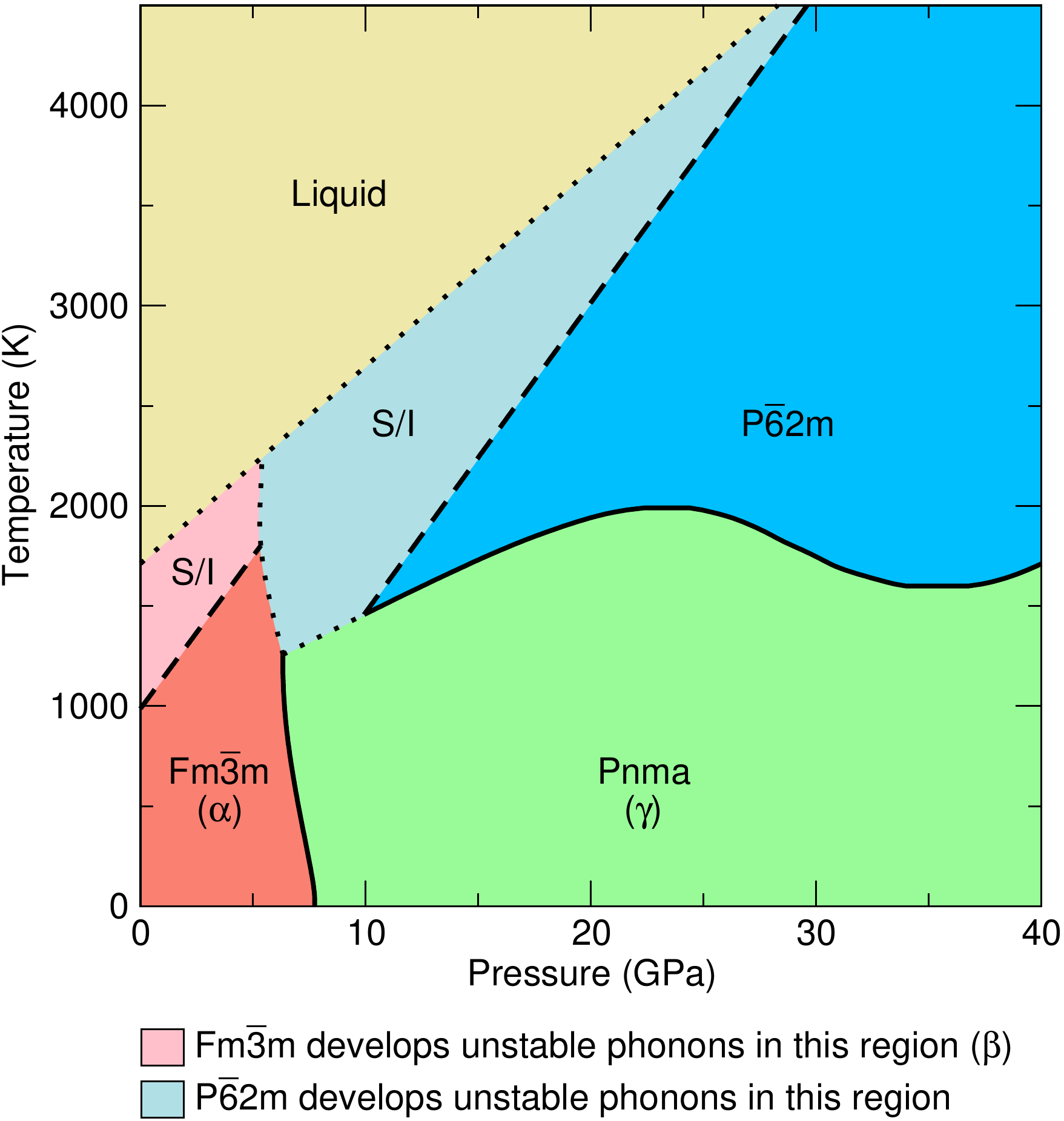}
  \caption{\label{fig:CaF_phasediagram} Calculated quasiharmonic phase
    diagram of CaF$_2$. We find the known $Fm\overline{3}m$ ($\alpha$)
    and $Pnma$ ($\gamma$) phases at low temperature, but our
    calculations indicate that a previously unreported phase of
    $P\overline{6}2m$ symmetry becomes stable at high temperature and
    pressure. `S/I' indicates the superionic region of the phase
    diagram, and dashed lines show boundaries due to unstable phonons
    at the quasiharmonic level.}
\end{figure}

The effects of nuclear zero-point motion and temperature are often
important and affect the relative stability of crystal phases,
particularly in cases where there are two or more structures lying
very close in energy \cite{Nelson-2015,H_Phase_IV_2012}. Given the
small enthalpy difference between the $P\overline{6}2m$ and $\gamma$
CaF$_2$ phases, we calculate the Gibbs free energy of these structures
in the QHA as a function of pressure, as well as that of
$Fm\overline{3}m$-CaF$_2$. Selecting the lowest Gibbs free energy
structure at each temperature and pressure gives the phase diagram
shown in Fig.~\ref{fig:CaF_phasediagram}. The solid-liquid phase
boundary (dotted black line) in Fig.~\ref{fig:CaF_phasediagram} is
taken from the work of Cazorla \textit{et al}
\cite{Cazorla-2015}. From this, we do indeed predict that the
$P\overline{6}2m$-CaF$_2$ structure is stabilised by temperature, in
the region $P\gtrsim 10$ GPa and $T\gtrsim 1500$ K. We remark here
that the exact phase boundaries in Fig.~\ref{fig:CaF_phasediagram} are
subject to some uncertainty depending on the choice of equation of
state used for the Gibbs free energy calculation. This uncertainty is
particularly noticeable along the $Pnma$ ($\gamma$)$-P\overline{6}2m$
phase boundary.

Both the $Fm\overline{3}m$ and $P\overline{6}2m$ structures develop
unstable phonon modes at sufficiently large volumes. For
$Fm\overline{3}m$-CaF$_2$, these first occur at a static-lattice
pressure between $-7$ and $-6$ GPa; for $P\overline{6}2m$ they first
occur between 0 and 1 GPa. Cazorla \textit{et al.} also report
unstable phonon modes for $Fm\overline{3}m$-CaF$_2$ above 4.5 GPa \cite{Cazorla-2015},
however we do not encounter any such instabilities in our
calculations. We find no unstable phonon modes in $\gamma$-CaF$_2$. In
Fig.~\ref{fig:CaF_phasediagram}, black dashed lines are used to divide
the regions of stability for the $Fm\overline{3}m$ and
$P\overline{6}2m$ in two. At temperatures below the lines, these
phases have volumes corresponding to stable phonons, while above the
lines they exhibit unstable phonon modes, and their Gibbs free
energies are extrapolations of quasiharmonic results. Dotted lines
separate regions where one or both phases have unstable phonon modes,
with the exception of the solid-liquid boundary.

The calculated volume coefficient of thermal expansion, $\alpha(P,T)$,
can be used to assess the validity of the QHA. Applying the criteria
of Karki \textit{et al.} \cite{Karki-2001} and Wentzcovitch \textit{et
  al.} \cite{Wentzcovitch-2004}, we expect the quasiharmonic
approximation to be accurate for
$T (\mbox{K}) \leq 28P\,(\mbox{GPa}) + 453$, with an uncertainty of
about 100 K. The lower half of the $\alpha$-$\gamma$ phase boundary,
and the $\gamma$-$P\overline{6}2m$ phase boundary near 40 GPa, are
therefore expected to be accurate within the QHA. At higher
temperatures, the QHA is expected to be less applicable as anharmonic
effects become increasingly important. Further information on the
calculation of $\alpha(P,T)$ and the validity of the QHA can be found
in the Appendix.

\subsection{Superionicity in CaF$_2$}

\begin{figure}
\centering
  \includegraphics[scale=0.46,clip]{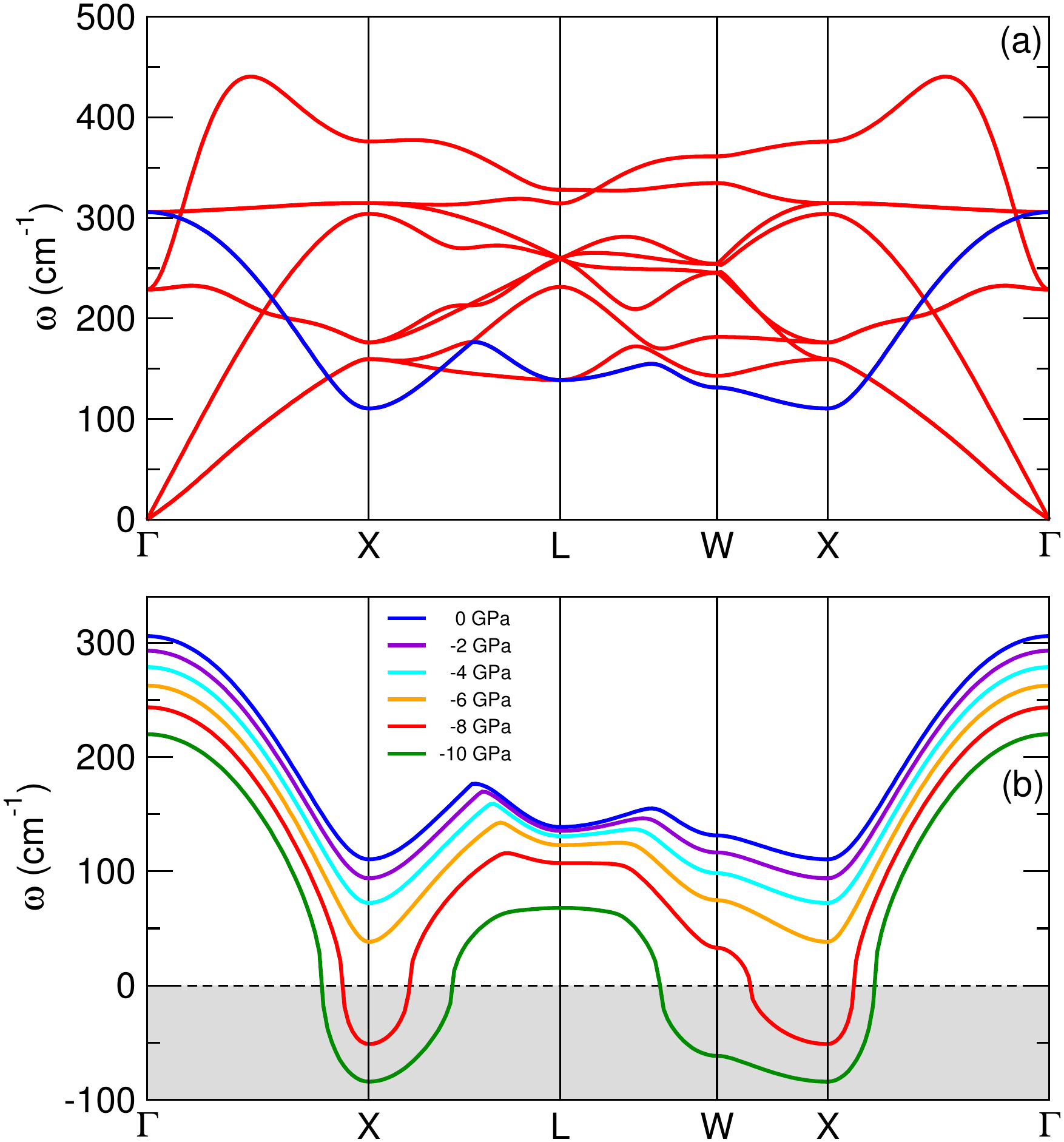}
  \caption{\label{fig:Fm3m_unstable} Softening of phonon modes in
    the fluorite $Fm\overline{3}m$-CaF$_2$ structure. (a) Phonon
    dispersion curves of this structure at a static-lattice pressure
    of 0 GPa (cell volume 41.79 \AA$^3$/f.u.). (b) The blue-coloured
    mode in (a) as a function of decreasing static pressure, from 0 to
    $-10$ GPa. The mode softens and first develops imaginary phonon
    frequencies at $X$ (shown as negative frequencies).}
\end{figure}

\begin{figure}
\centering
  \includegraphics[scale=0.46,clip]{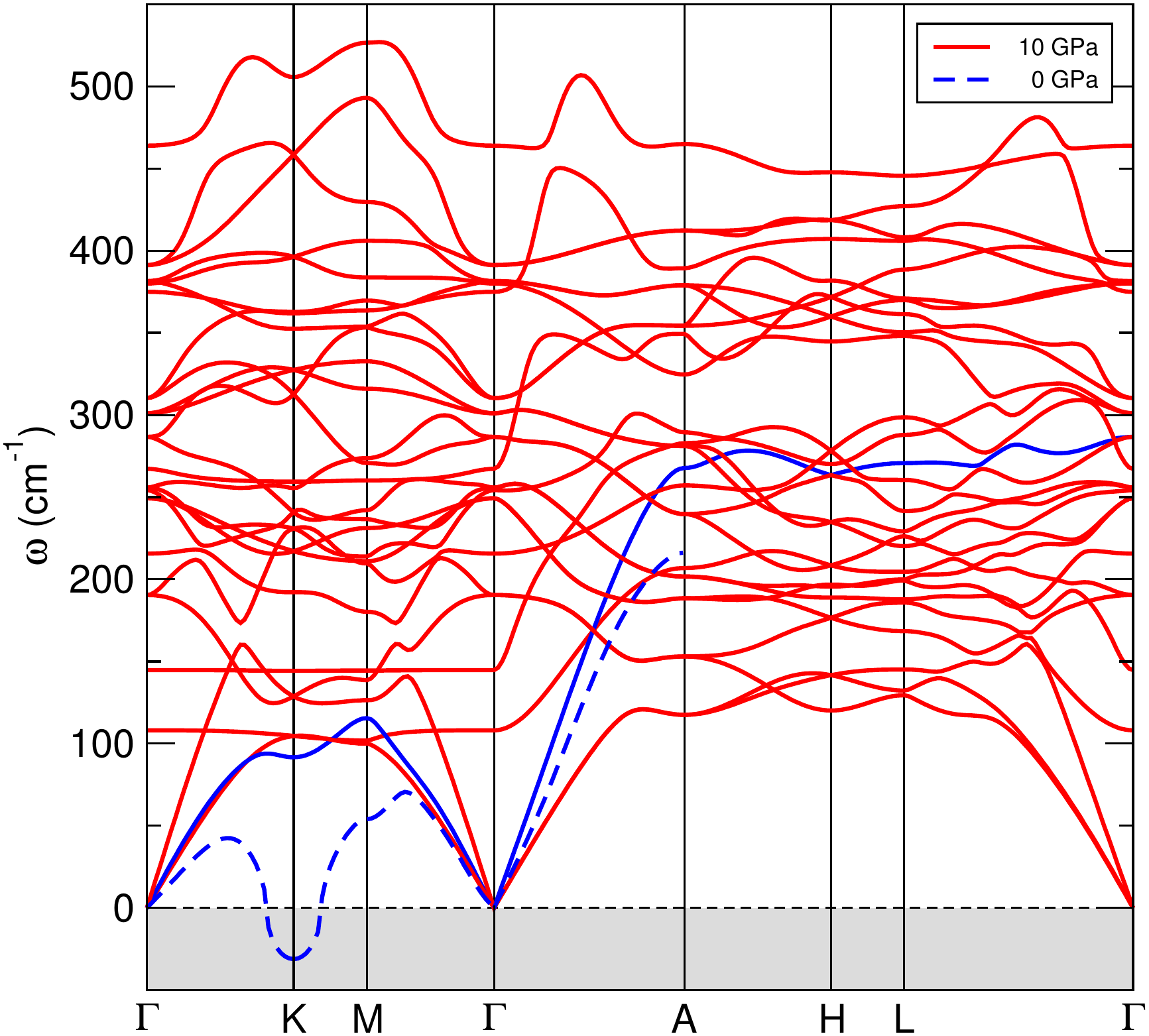}
  \caption{\label{fig:P62m_unstable} Phonon dispersion curves for
    the proposed $P\overline{6}2m$-CaF$_2$ structure. This phase has
    stable phonons at a static pressure of 10 GPa (cell volume 34.45
    \AA$^3$/f.u.). The mode coloured blue softens and becomes unstable
    at the Brillouin zone $K$ point with decreasing pressure, as shown
    by the blue dashed curve. Only the portion from $\Gamma$ to $A$ is
    shown at 0 GPa. }
\end{figure}

The onset of superionicity in $\beta$-CaF$_2$ has been discussed in
connection with the formation of unstable phonon modes in the fluorite
CaF$_2$ structure \cite{Boyer-1980}. Indeed, this is the criterion we
have used in Fig.~\ref{fig:CaF_phasediagram}, where we label the
region where $Fm\overline{3}m$-CaF$_2$ has unstable phonons as
superionic (`S/I'), or $\beta$. By this criterion, the
$\alpha$-$\beta$ transition is calculated to occur at $\approx1000$ K
at 0 GPa, which is actually in rough agreement with the experimentally
observed transition temperature of 1400 K, considering that the QHA
should be inaccurate near unstable phonon
modes. Fig.~\ref{fig:Fm3m_unstable}(a) shows the phonon dispersion
relations in $Fm\overline{3}m$-CaF$_2$ at a static-lattice pressure of
0 GPa, while Fig.~\ref{fig:Fm3m_unstable}(b) shows how unstable modes
develop in this structure with increasing volume (decreasing
static-lattice pressure). Unstable phonon modes are first encountered
at the Brillouin zone $X$ point. The corresponding atomic
displacements in this unstable mode leave Ca$^{2+}$ ions fixed, while
F$^{-}$ ions are displaced along the [100], [010] or [001] directions
(referred to a conventional cubic cell for
$Fm\overline{3}m$-CaF$_2$). The connection to superionicity is that
these directions also correspond to easy directions for F$^{-}$ ion
diffusion in the fluorite structure, and are almost barrierless at
volumes corresponding to unstable $X$ phonons \cite{Gupta-2012}.

Fig.~\ref{fig:Fm3m_unstable}(b) also shows that at even larger
volumes, unstable phonon modes develop at the Brillouin zone $W$ point
and that the entire $W$$-$$X$
branch becomes soft. As is the case at $X$, the corresponding phonon
modes involve only F$^{-}$ ion displacements, though in different
directions: at $W$, F$^{-}$ ions displace along the [011] and [0-11]
directions. This may explain the observed gradual onset of
superionicity in the fluorite structure \cite{Hull-2004}: as volume
increases, further low-energy diffusion pathways corresponding to
unstable phonon modes are opened up in the lattice.

\begin{figure}
\centering
  \includegraphics[scale=0.15,clip]{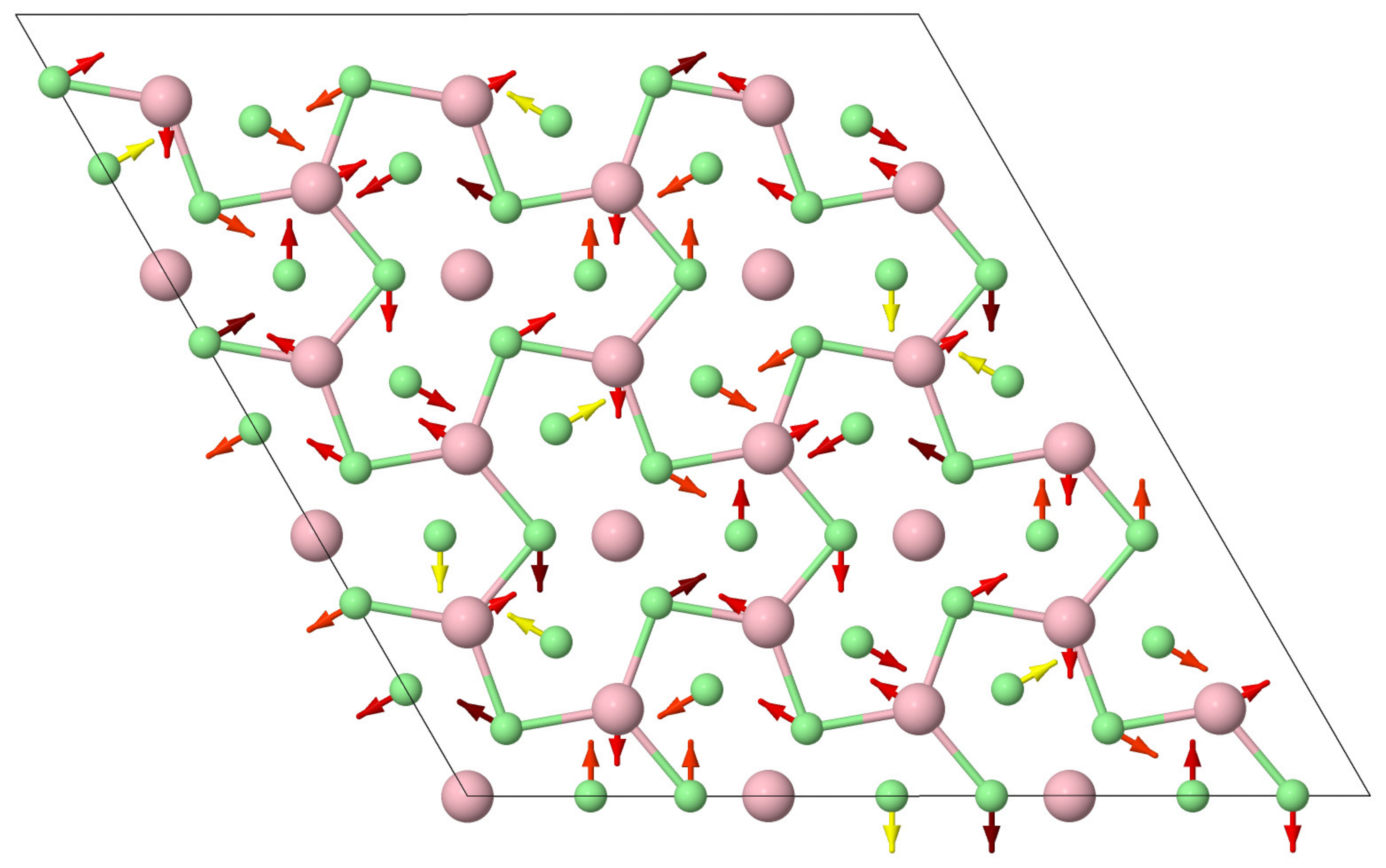}
  \caption{\label{fig:Unstable_P62m_K} Illustration of the unstable
    phonon mode in $P\overline{6}2m$-CaF$_2$ at the Brillouin zone $K$
    point, shown here at 0 GPa. A 3$\times$3$\times$1 slab of the
    structure is depicted, viewed along the $c$-axis as in the
    left-hand panel of Fig.~\ref{fig:P62m_and_Pnma}(a). Arrows
    indicate the direction of movement of atoms in this mode, and are
    colour-coded according to their relative amplitudes: yellow for
    largest amplitude, through to dark red for smallest
    amplitude. Calcium atoms are in red, fluorine atoms in
    green.}
\end{figure}

Fig.~\ref{fig:P62m_unstable} shows the phonon dispersion relations in
$P\overline{6}2m$. With increasing volume, this structure first
develops unstable phonon modes at the Brillouin zone $K$ point, and
the atomic displacements of Ca$^{2+}$ and F$^{-}$ ions in the
corresponding mode are depicted in
Fig.~\ref{fig:Unstable_P62m_K}. This mode is similar to the unstable
mode found in fluorite CaF$_2$ at $X$, in the sense that it involves
displacements of F$^{-}$ ions and a sublattice of Ca$^{2+}$ ions which
remain fixed. F$^{-}$ ions move along the [120], [210] or [1-10]
directions, and all displacements are confined to the $ab$-plane
only. The $P\overline{6}2m$ structure can be visualised as layer-like:
in Fig.~\ref{fig:Unstable_P62m_K}, all atoms that are linked by bonds
belong to the same layer, and all `isolated' atoms belong to a
different layer. We find that Ca$^{2+}$ ions in alternating layers
remain fixed in the unstable phonon mode. By analogy with fluorite
CaF$_2$, we propose that $P\overline{6}2m$-CaF$_2$ also undergoes a
superionic phase transition accompanying this phonon mode, and we
label the region where $P\overline{6}2m$ has unstable phonon modes in
Fig.~\ref{fig:CaF_phasediagram} as superionic (`S/I'). As is the case
for the fluorite structure, it is possible that other compounds in the
$P\overline{6}2m$ structure could also exhibit superionicity.

The pressure-temperature phase diagram of CaF$_2$ has recently been
examined by Cazorla \textit{et al}.~\cite{Cazorla-2013,Cazorla-2015}. In
addition to the known $\alpha$, $\beta$ and $\gamma$ phases, the
authors propose a high-temperature phase transition from $\gamma$ to a
new $\delta$ phase, which in turn is predicted to undergo a superionic
transition at even higher temperatures, to a phase labelled
$\epsilon$-CaF$_2$. A $P2_1/c$ symmetry structure was proposed for the
$\delta$-phase \cite{Cazorla-2015}, however we find that this structure
is close to $Pnma$ symmetry, and relaxing it using DFT gives the
$\gamma$-CaF$_2$ structure. The phase diagram of
Fig.~\ref{fig:CaF_phasediagram} is in qualitative agreement with these
results, where we identify the $\delta$ phase with our predicted
$P\overline{6}2m$ structure, and the $\epsilon$ superionic phase with
the region where $P\overline{6}2m$ has unstable phonon modes.

\subsection{Optical bandgaps in CaF$_2$}
Our calculations show that optical bandgaps in CaF$_2$ initially
increase but then remain relatively constant over the pressure range
0$-$70 GPa (Fig.~\ref{fig:optical_bandgaps}). We also show, using a
dashed line Fig.~\ref{fig:optical_bandgaps}, the calculated bandgap
for the $P\overline{6}2m$-CaF$_2$ phase, which begins to slowly
decrease above 50 GPa. Semilocal DFT calculations using GGA
functionals have also shown that this occurs for $\gamma$-CaF$_2$
above 70 GPa \cite{Shi-2009}. The optical bandgap for $\gamma$-CaF$_2$
lies \mbox{0.9$-$1.2 eV} above that of $P\overline{6}2m$-CaF$_2$ in
the pressure range 10$-$70 GPa, suggesting that the formation of
$P\overline{6}2m$-CaF$_2$ might be detectable in optical
measurements. We also give the electronic DOS of the $P\overline{6}2m$
phase in the Appendix.

Low temperature CaF$_2$ has been proposed as an internal pressure
standard \cite{Gerward-1992}. Our calculations of the bandgap shows
that it remains a wide-gap insulator up to at least 70 GPa, and likely
retains its superior optical properties up until high pressures.

\section{Conclusions}
We have explored Be-, Mg- and CaF$_2$ at pressures up to 70 GPa
through DFT calculations and computational structure
searching.

BeF$_2$ has a large number of polymorphs at ambient pressures, and
shares many of these with SiO$_2$, such as $\alpha,\beta$-quartz and
$\alpha,\beta$-cristobalite. Our searches show that BeF$_2$ has
open-framework zeolite-like polymorphs, and that framework structures
predicted in BeF$_2$ are energetically relevant to SiO$_2$,
highlighting the utility of structure searching in model systems. At
higher pressures, we find that BeF$_2$ is stable in the moganite
structure between 11.6 and 30.1 GPa, and stable in the CaCl$_2$
structure between 30.1 and 57.5 GPa.

In MgF$_2$, we find that the $Pbca$-symmetry `O-I' TiO$_2$ structure
is a stable intermediary between the pyrite and cotunnite MgF$_2$
phases, and is the lowest enthalpy MgF$_2$ structure between 39.6 and
44.1 GPa. A class of polymorphs for SiO$_2$ intermediate to the
CaCl$_2$ and $\alpha$-PbO$_2$ structures, which are relevant at Earth
mantle pressures and are close to stability near 100 GPa in SiO$_2$,
also occur in MgF$_2$ but at much lower pressures ($\approx10$ GPa).

We find that the Fe$_2$P-type $P\overline{6}2m$-symmetry structure for
CaF$_2$ lies close in enthalpy to the known $\gamma$-CaF$_2$ phase
over the pressure range 0$-$70 GPa. Calculations using the QHA show
that this structure is stabilised at high pressure and temperature
($P\gtrsim 10$ GPa and $T\gtrsim 1500$ K). The $P\overline{6}2m$
structure develops unstable phonon modes at high temperatures, which
we propose is associated with a superionic transition in this
structure. $P\overline{6}2m$-CaF$_2$ and its region of phonon
instability are consistent with the recently proposed $\delta$ and
$\epsilon$ CaF$_2$ phases.

Be-, Mg- and CaF$_2$ are wide-gap insulators. Calculations using the
HSE06 functional show that the bandgaps in BeF$_2$ and MgF$_2$ are
tunable with pressure, rising by 0.06 eV/GPa and 0.04 GPa over the
pressure range 0$-$70 GPa. The optical bandgaps in CaF$_2$ are instead
relatively constant over this pressure range.

\section{Acknowledgements}
R. J. N. acknowledges financial support from the Engineering and
Physical Sciences Research Council (EPSRC) of the U.K. [EP/J017639/1].
C. J. P. acknowledges financial support from EPSRC [EP/G007489/2], a
Leadership Fellowship [EP/K013688/1] and a Royal Society Wolfson
research merit award.  J. R. N., R. J. N. and C. J. P. acknowledge use
of the Archer facilities of the U.K.'s national high-performance
computing service, for which access was obtained via the UKCP
consortium [EP/K013688/1], [EP/K014560/1]. J. R. N. acknowledges the
support of the Cambridge Commonwealth Trust.

\pagebreak
\widetext

\large 

\centerline{\textbf{Appendix for:}}

\vspace{0.1cm}

\centerline{\textbf{``High-pressure phases of group II difluorides:
    polymorphism and superionicity''}}

\vspace{0.1cm}

\centerline{\textbf{J.~R.~Nelson,$^{1,*}$ R.~J.~Needs,$^1$ and C.~J.~Pickard$^{2,3}$}}

\vspace{0.1cm}

\normalsize
\centerline{\textit{$^1$Theory of Condensed Matter Group, Cavendish Laboratory,}}

\centerline{\textit{J.~J.~Thomson Avenue, Cambridge CB3 0HE, United Kingdom}}

\centerline{\textit{$^2$Department of Materials Science and Metallurgy, University of
Cambridge,}}

\centerline{\textit{27 Charles Babbage Road, Cambridge CB3 0FS, United Kingdom}}

\centerline{\textit{$^3$Advanced Institute for Materials Research, Tohoku University, }}

\centerline{\textit{2-1-1 Katahira, Aoba, Sendai, 980-8577, Japan}}

\vspace{0.1cm}

\centerline{$^{*}$Email: \texttt{jn336@cam.ac.uk}}


\setcounter{section}{0}
\setcounter{page}{1}
\makeatletter

\section{Electronic structure calculations} \label{sec:esc} 

All calculations are carried out with version 8.0 of the
\textsc{Castep} plane$-$wave pseudopotential DFT code.

\vspace{0.2cm}

Our calculations of the enthalpy vs.~pressure curves in the main paper
and in this appendix, lattice parameters, and supercell phonon
calculations all use default ultrasoft pseudopotentials for Be, Mg,
Ca, F, Si and O as generated internally by the \textsc{Castep} code
and the Perdew$-$Burke$-$Ernzerhof (PBE) exchange$-$correlation
functional. A plane$-$wave cutoff of 800 eV, Brillouin zone sampling
density of at least $2\pi\times 0.04$ \AA$^{-1}$, grid scale of 2 (for
representing the density) and fine grid scale of 2.5 (grid for
ultrasoft core or augmentation charges) are used in these
calculations. Calculations of phonon frequencies use the
finite$-$displacement supercell method in \textsc{Castep}, and the
cumulant force constant matrix cutoff scheme
\cite{Parlinski-1997,Ye-2004}. Details of the supercell sizes used in
finite$-$displacement phonon calculations are given in subsequent
sections.

\vspace{0.2cm}

Bandgap calculations (Fig.~3 of the main paper) are instead carried
out with norm$-$conserving pseudopotentials, which are required in
order to use hybrid density functionals in \textsc{Castep}. Generation
strings for these potentials are taken from the version 16.0
\textsc{Castep} Norm-Conserving Pseudopotential (`NCP') library, and
used in version 8.0 of the code. Our bandgap calculations use the
`HSE06' variant of the Heyd-Scuseria-Ernzerhof hybrid functional, a
plane$-$wave cutoff of 1600 eV, a Brillouin zone sampling density of
at least $2\pi\times 0.1$ \AA$^{-1}$, and grid and fine grid scales of
2 and 2.5 respectively. The structures for bandgap calculations were
obtained by geometry optimisation using the same plane$-$wave cutoff,
Brillouin zone sampling and grid scales, but with the PBE functional
instead of HSE06.

\vspace{0.2cm}

Optical bandgaps for BeF$_2$ are calculated for the $\alpha$-quartz,
coesite-I, moganite, CaCl$_2$ and $\alpha$-PbO$_2$ structures and
shown in Fig.~3 of the main paper, over the pressure ranges for which
these structures are calculated to be stable at the static-lattice
level. We therefore exclude the small stability range of
$\alpha$-cristobalite BeF$_2$ (0$-$0.4 GPa), and use the coesite-I
structure instead of coesite-II on the grounds that these two
structures are almost identical at low pressures. Optical bandgaps for
MgF$_2$ are calculated for the rutile, $\alpha$-PbO$_2$, pyrite, O-I
and cotunnite structures and likewise shown in Fig.~3 of the main
paper. Finally for CaF$_2$, we show optical bandgaps for the fluorite,
cotunnite and $P\overline{6}2m$ structures.

\newpage

\section{Structure data}

\subsection{Beryllium fluoride structure data}

\textbf{Moganite, space group $C2/c$ (\#15)} \\
The table below gives the DFT-relaxed BeF$_2$ moganite structure at 20 GPa. A conventional cell \mbox{(36 atoms)} with full Hermann-Mauguin (HM) symbol $C12/c1$ (standard setting for this space group) is given; the primitive cell has 18 atoms. \\
Fig.~2(a) of the main paper illustrates this structure, but uses
different lattice vectors, obtained from those below through the
relations $\textbf{a}' = \textbf{a} + \textbf{c}$,
$\textbf{b}' =-\textbf{b}$, $\textbf{c}' = -\textbf{c}$. This has full
HM symbol $I12/c1$ and a smaller monoclinic angle
($\beta = 90.446^{\circ}$).

\begin{center}
\begin{tabular}{llllcccc} \hline
\multicolumn{3}{c}{Lattice parameters}& & \multicolumn{3}{c}{Atomic coordinates} & Wyckoff  \\
 &(\AA, deg.)& &Atom&$x$&$y$&$z$& site \\ \hline
$a$=12.008    &$b$=4.191    &$c$=7.103            &Be&0.0000&0.9572&0.2500&4e \\
$\alpha$=90.000 &$\beta$=125.818 &$\gamma$=90.000 &Be&0.1658&0.3268&0.1989&8f \\
 &  &  &F&0.2909&0.1705&0.2289&8f \\
 &  &  &F&0.1237&0.1811&0.3455&8f \\
 &  &  &F&0.0421&0.2694&0.9492&8f \\ \hline
\end{tabular}
\end{center}

\vspace{0.3cm}

\textbf{CaCl$_2$ structure, space group $Pnnm$ (\#58)} \\
The unit cell is primitive and has $Z$=2 (6 atoms). The positions
below correspond to 50 GPa, and the structure is depicted in Fig.~2(b)
of the main paper.

\begin{center}
\begin{tabular}{llllcccc} \hline
\multicolumn{3}{c}{Lattice parameters}& & \multicolumn{3}{c}{Atomic coordinates} & Wyckoff  \\
 &(\AA, deg.)& &Atom&$x$&$y$&$z$& site \\ \hline
$a$=3.796    &$b$=3.959    &$c$=2.445            &Be&0.0000&0.0000&0.0000&2a \\
$\alpha$=90.000 &$\beta$=90.000 &$\gamma$=90.000 &F&0.2736&0.3233&0.0000&4g \\ \hline
\end{tabular}
\end{center}

\vspace{0.4cm}

\subsection{Magnesium fluoride structure data}
\textbf{Orthorhombic-I structure, space group $Pbca$ (\#61)} \\
The unit cell is primitive and has $Z$=8 (24 atoms). The positions
below correspond to 42 GPa, and the structure is depicted in Fig.~5 of
the main paper.

\begin{center}
\begin{tabular}{llllcccc} \hline
\multicolumn{3}{c}{Lattice parameters}& & \multicolumn{3}{c}{Atomic coordinates} & Wyckoff  \\
 &(\AA, deg.)& &Atom&$x$&$y$&$z$& site \\ \hline
$a$=4.782    &$b$=4.556    &$c$=8.863            &Mg&0.9586&0.2259&0.8870&8c \\
$\alpha$=90.000 &$\beta$=90.000 &$\gamma$=90.000 &F&0.1099&0.1721&0.2008&8c \\ 
 & & &F&0.2339&0.9977&0.4653&8c \\ \hline
\end{tabular}
\end{center}

\vspace{0.4cm}

\subsection{Calcium fluoride structure data}
\textbf{Fe$_2$P-type structure, space group $P\overline{6}2m$ (\#189)} \\
The unit cell is primitive and has $Z$=3 (9 atoms). The positions
below correspond to 30 GPa, and the structure is depicted in Fig.~7(a)
of the main paper.

\begin{center}
\begin{tabular}{llllcccc} \hline
\multicolumn{3}{c}{Lattice parameters}& & \multicolumn{3}{c}{Atomic coordinates} & Wyckoff  \\
 &(\AA, deg.)& &Atom&$x$&$y$&$z$& site \\ \hline
$a$=5.697    &$b$=5.697    &$c$=3.255            &Ca&0.0000&0.0000&0.5000&1b \\
$\alpha$=90.000 &$\beta$=90.000 &$\gamma$=120.000&Ca&0.3333&0.6666&0.0000&2c \\ 
 & & &F&0.7433&0.0000&0.0000&3f \\
 & & &F&0.4126&0.0000&0.5000&3g \\ \hline
\end{tabular}
\end{center}

\subsection{Low$-$pressure SiO$_2$ polymorphs}
Three low$-$enthalpy structures for BeF$_2$, labelled $P2_12_12_1$-I, $C2/c-4\times 2$, and $P2_1/c$ in Fig.~1(a) of the main paper, are examined as possible low$-$enthalpy polymorphs for SiO$_2$. The monoclinic angle in $P2_1/c$ is close to 90$^{\circ}$ ($=$89.958$^{\circ}$), and so we also examine a higher symmetry version of this with $\alpha$=$\beta$=$\gamma$=90$^{\circ}$, with the space group $Pnma$. Structure data for these polymorphs are shown below, alongside their enthalpies relative to the $\alpha$-quartz phase. All data is given at 0 GPa. \\

\textbf{$C2/c-4\times 2$ structure, space group \#15 ($-$15.7 meV/f.u.)} \\  
A conventional unit cell has $Z$=8 (24 atoms), and is given below. This has full HM symbol $C12/c1$.

\begin{center}
\begin{tabular}{llllcccc} \hline
\multicolumn{3}{c}{Lattice parameters}& & \multicolumn{3}{c}{Atomic coordinates} & Wyckoff  \\
 &(\AA, deg.)& &Atom&$x$&$y$&$z$& site \\ \hline
$a$=8.940    &$b$=5.031    &$c$=8.977             &Si&0.1877&0.1694&0.3129&8f \\
$\alpha$=90.000 &$\beta$=111.563 &$\gamma$=90.000 &O &0.2055&0.5806&0.7058&8f \\ 
 & & &O&0.0000&0.0798&0.2500&4e \\
 & & &O&0.2500&0.2500&0.5000&4d \\ \hline
\end{tabular}
\end{center}

\vspace{0.3cm}

\textbf{$P2_1/c$ structure, space group \#14 ($-$5.9 meV/f.u.)} \\  
The unit cell is primitive and has $Z$=8 (24 atoms). The cell given below has full HM symbol $P12_1/n1$ (ITA unique axis $b$, cell choice 2).

\begin{center}
\begin{tabular}{llllcccc} \hline
\multicolumn{3}{c}{Lattice parameters}& & \multicolumn{3}{c}{Atomic coordinates} & Wyckoff  \\
 &(\AA, deg.)& &Atom&$x$&$y$&$z$& site \\ \hline
$a$=8.583    &$b$=9.626    &$c$=5.207            &Si&0.0618&0.1781&0.2632&4e \\
$\alpha$=90.000 &$\beta$=89.958 &$\gamma$=90.000 &Si&0.3654&0.3576&0.2649&4e \\ 
 & & &O&0.1448&0.0263&0.2455&4e \\
 & & &O&0.1914&0.2978&0.2191&4e \\
 & & &O&0.4278&0.3066&0.5443&4e \\
 & & &O&0.4830&0.3001&0.0441&4e \\ \hline
\end{tabular}
\end{center}

\vspace{0.3cm}

\textbf{$P2_12_12_1$-I structure, space group \#19 ($-$5.5 meV/f.u.)} \\
The unit cell is primitive and has $Z$=8 (24 atoms).

\begin{center}
\begin{tabular}{llllcccc} \hline
\multicolumn{3}{c}{Lattice parameters}& & \multicolumn{3}{c}{Atomic coordinates} & Wyckoff  \\
 &(\AA, deg.)& &Atom&$x$&$y$&$z$& site \\ \hline
$a$=5.195    &$b$=8.519    &$c$=9.560            &Si&0.2933&0.6088&0.8607&4a \\
$\alpha$=90.000 &$\beta$=90.000 &$\gamma$=90.000 &Si&0.7068&0.1942&0.3192&4a \\ 
 & & &O&0.7237&0.1042&0.4693&4a \\
 & & &O&0.9249&0.3308&0.3101&4a \\
 & & &O&0.5746&0.7739&0.2020&4a \\
 & & &O&0.2499&0.5688&0.3056&4a \\ \hline
\end{tabular}
\end{center}

\vspace{0.3cm}

\textbf{$Pnma$ structure, space group \#62 ($-$2.2 meV/f.u.)} \\
The unit cell is primitive and has $Z$=8 (24 atoms).

\begin{center}
\begin{tabular}{llllcccc} \hline
\multicolumn{3}{c}{Lattice parameters}& & \multicolumn{3}{c}{Atomic coordinates} & Wyckoff  \\
 &(\AA, deg.)& &Atom&$x$&$y$&$z$& site \\ \hline
$a$=9.415    &$b$=5.285    &$c$=8.683            &Si&0.6850&0.2500&0.5529&4c \\
$\alpha$=90.000 &$\beta$=90.000 &$\gamma$=90.000 &Si&0.3577&0.2500&0.6389&4c \\ 
 & & &O&0.5303&0.2500&0.6363&4c \\
 & & &O&0.3087&0.2500&0.8176&4c \\
 & & &O&0.2041&0.4997&0.0538&8d \\ \hline
\end{tabular}
\end{center}

\newpage

\section{Convex hulls for Be, Mg and Ca$-$F at 60 GPa}

In addition to our BeF$_2$, MgF$_2$ and CaF$_2$ searches, we carry out variable stoichiometry structure searches for the Be-F, Mg-F and Ca-F systems at 60 GPa. Convex hulls depicting the results of these searches are shown below.

\begin{figure*}[htp]
  \centering
  \subfigure{\includegraphics[scale=0.28,clip]{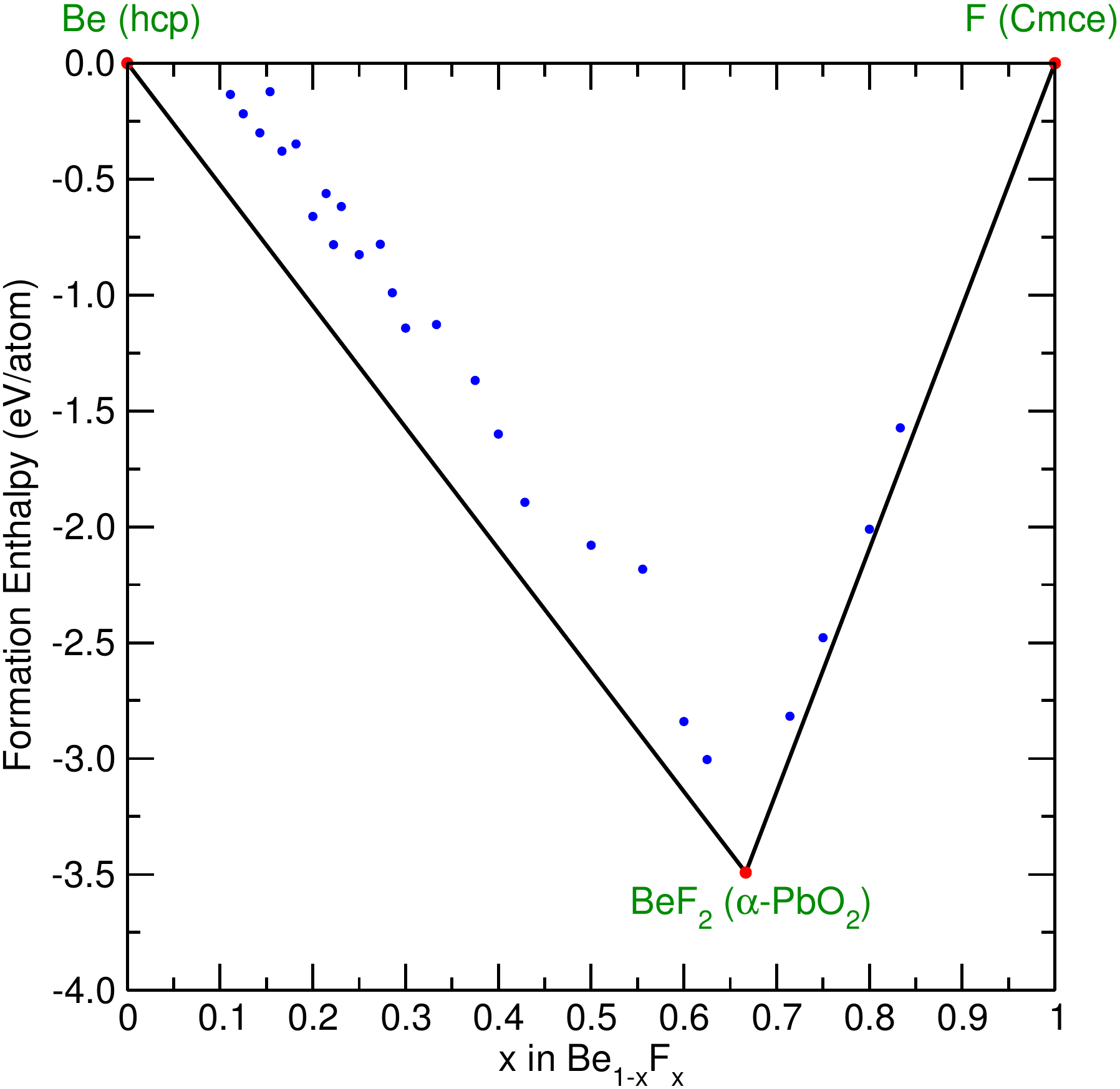}}\quad
  \subfigure{\includegraphics[scale=0.28,clip]{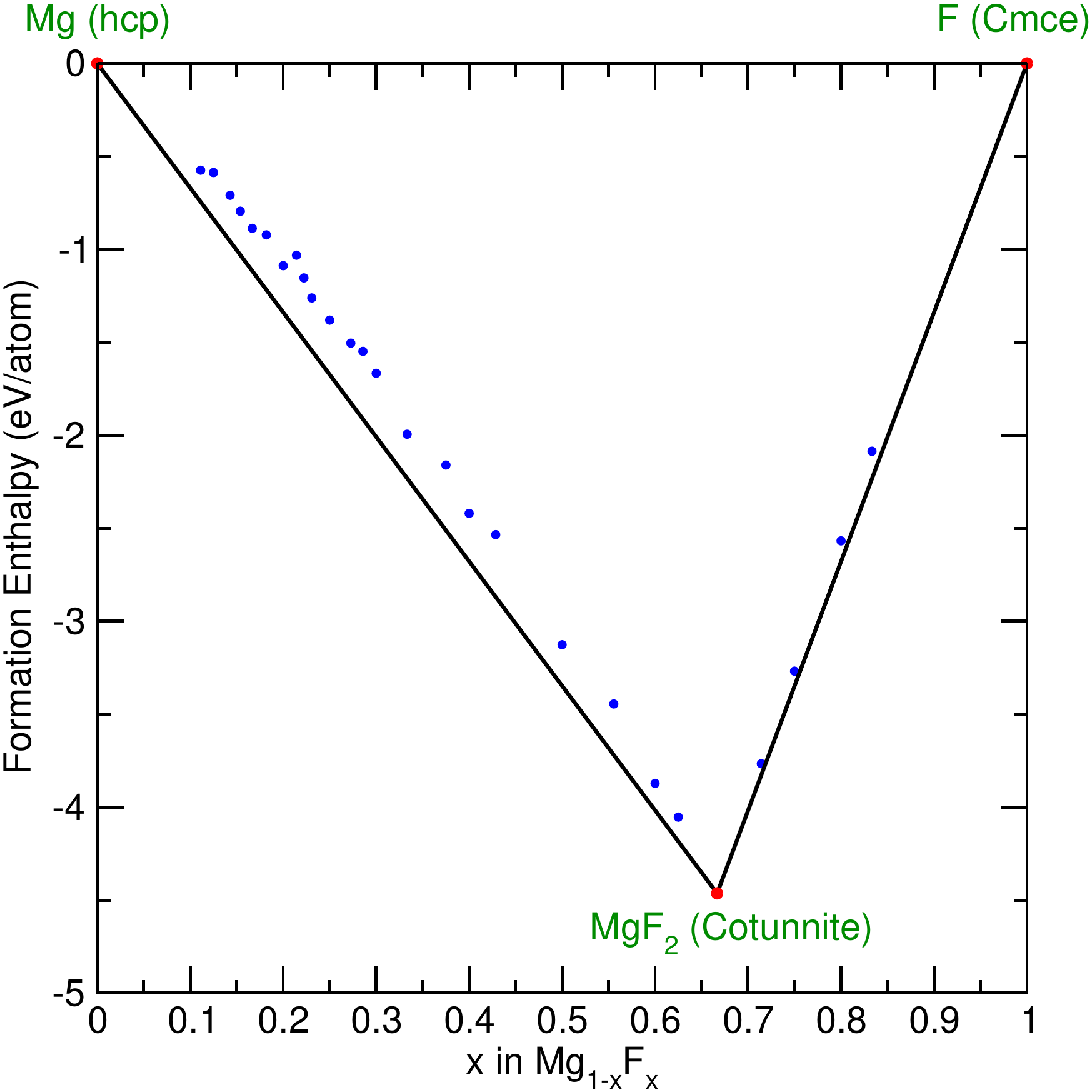}}\quad
  \subfigure{\includegraphics[scale=0.28,clip]{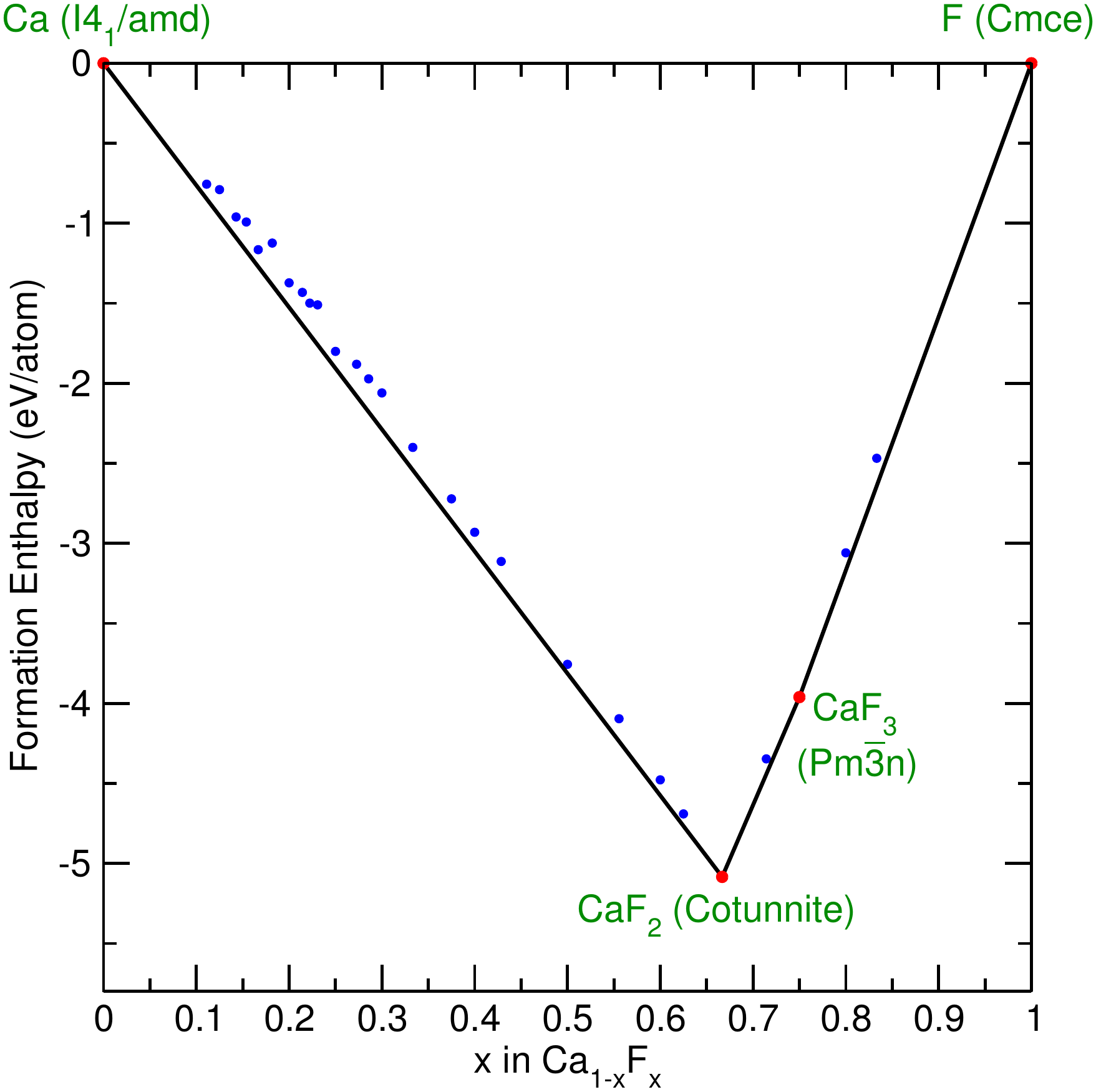}}
  \caption{\label{fig:Hulls} Convex hulls showing the results of structure searching at the static-lattice level of theory at 60 GPa in the Be-F (left), Mg-F (middle) and Ca-F (right) systems.}
\end{figure*}

In each case, we search over 29 stoichiometries, including the elements (Be or Mg or Ca, and F). The convex hulls depict the lowest enthalpy structure at each stoichiometry, relaxed to the same standard as the results in the main paper (800 eV basis set cutoff, Brillouin zone sampling of 2$\pi$$\times$0.04 \AA$^{-1}$, PBE exchange).

\vspace{0.2cm}

At 60 GPa, beryllium has the $\alpha$-Be hexagonal closed-packed (hcp) structure, space group $P6_3/mmc$ \cite{Lazicki-2012}, which we also verify directly through structure searches.  We find a molecular crystal comprised of F$_2$ molecules with space group $Cmce$ to be the lowest enthalpy fluorine structure at 60 GPa; this also applies to the Mg-F and Ca-F systems. Of the stoichiometries searched, only BeF$_2$ is stable at this pressure in the Be-F system.

\vspace{0.2cm}

Magnesium undergoes a phase transition from hcp to bcc between about 45 and 60 GPa; within this pressure range, the coexistence of the hcp and bcc phases is observed at low temperatures \cite{Stinton-2014}. Our PBE$-$DFT calculations find that the hcp$-$Mg phase is very slightly lower in enthalpy than bcc$-$Mg at 60 GPa, so we use the hcp$-$Mg phase in our hulls above. As is the case for BeF$_2$, we find that MgF$_2$ is the only stable stoichiometry in the Mg-F system at 60 GPa.

\vspace{0.2cm}

DFT and structure searching calculations predict that the $\beta$-tin $I4_1/amd$-symmetry structure is the lowest enthalpy phase for calcium at 60 GPa \cite{Oganov-2010}, and we also confirm this in the present study. The $\beta$-tin phase is well known to be slightly at odds with experimental results, which instead observe simple cubic calcium (or distorted versions of it) at 60 GPa \cite{Li-2012}, although the $\beta$-tin phase has been synthesised near 35 GPa. We carry out our calculations with the lowest enthalpy DFT phase for calcium ($\beta$-tin) in the Ca$-$F hulls above. Our searches find that both CaF$_2$ and CaF$_3$ are stable stoichiometries at 60 GPa.

\newpage

\section{Electronic and phonon DOS of selected structures}
\begin{figure*}[htp]
  \centering
  \subfigure{\includegraphics[scale=0.38,clip]{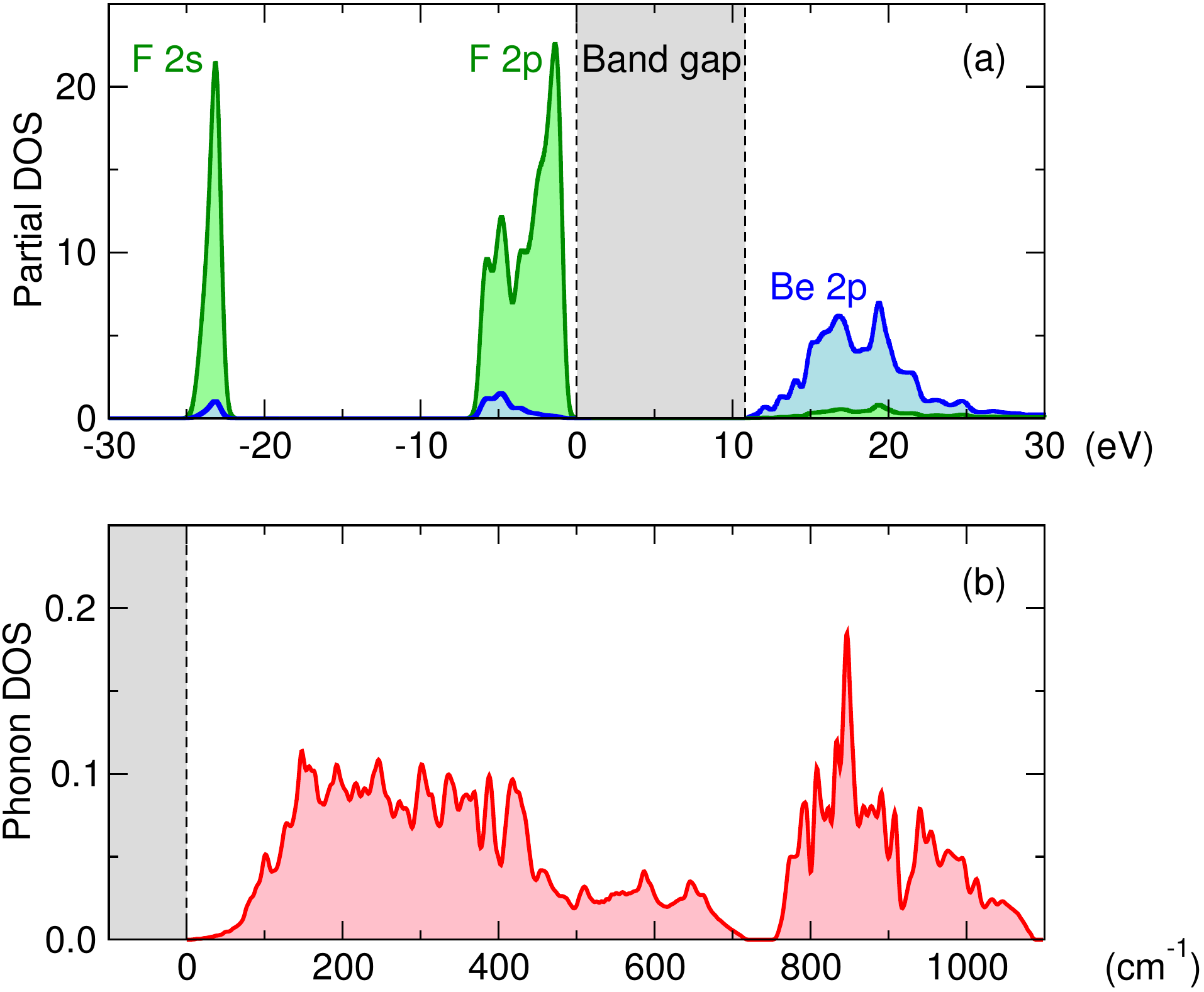}}\quad
  \subfigure{\includegraphics[scale=0.38,clip]{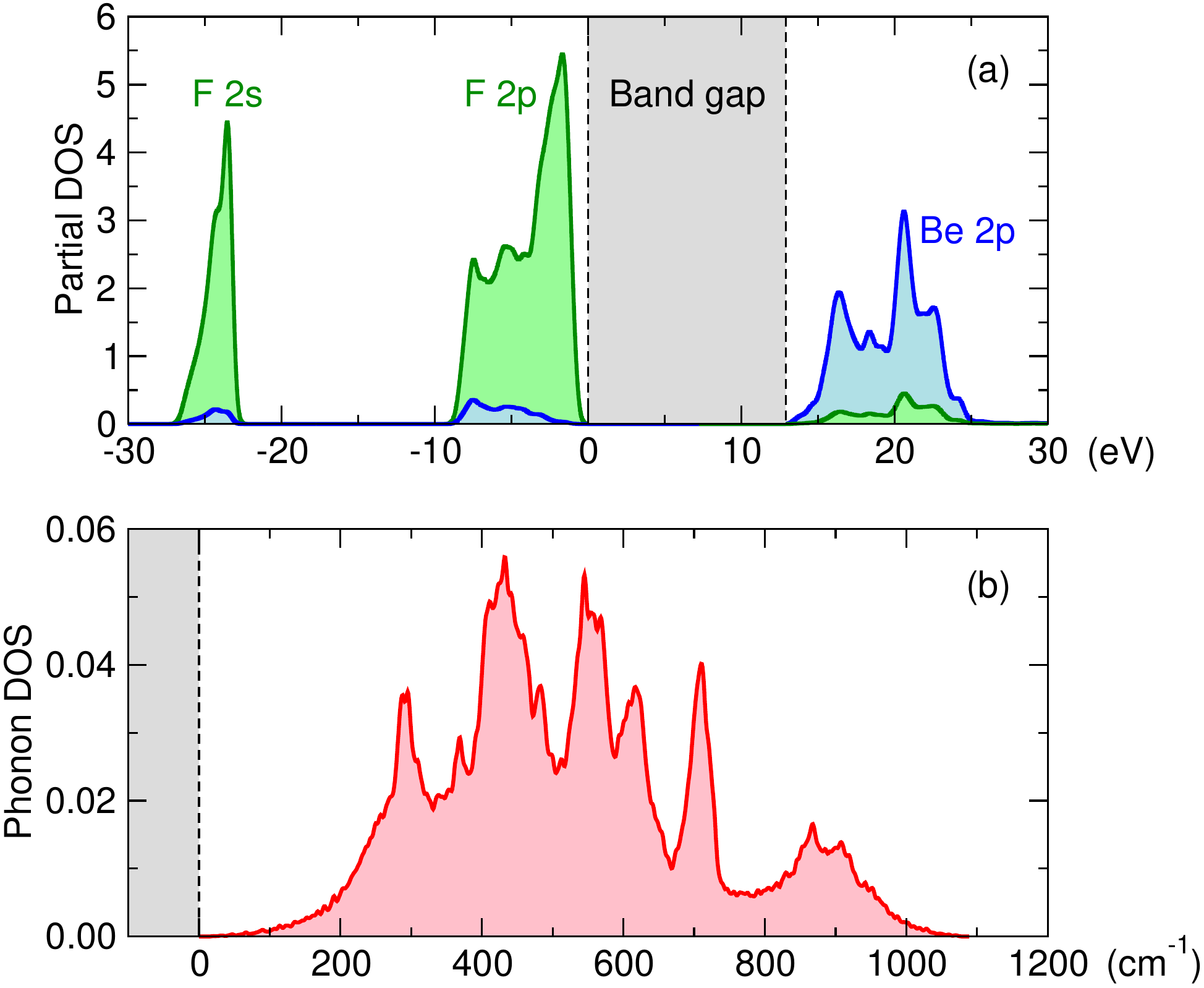}}
  \caption{\label{fig:moganite_and_cacl2} Electronic (a) and phonon (b) density of states for the moganite$-$BeF$_2$ phase at 20 GPa (left panel), and for the CaCl$_2$$-$BeF$_2$ phase at 50 GPa (right panel). A 4$\times$4$\times$3 (864 atom) and 4$\times$4$\times$6 (576 atom) supercell is used for the moganite and CaCl$_2$ phonon calculations, respectively.}
\end{figure*}

\begin{figure*}[htp]
  \centering
  \subfigure{\includegraphics[scale=0.38,clip]{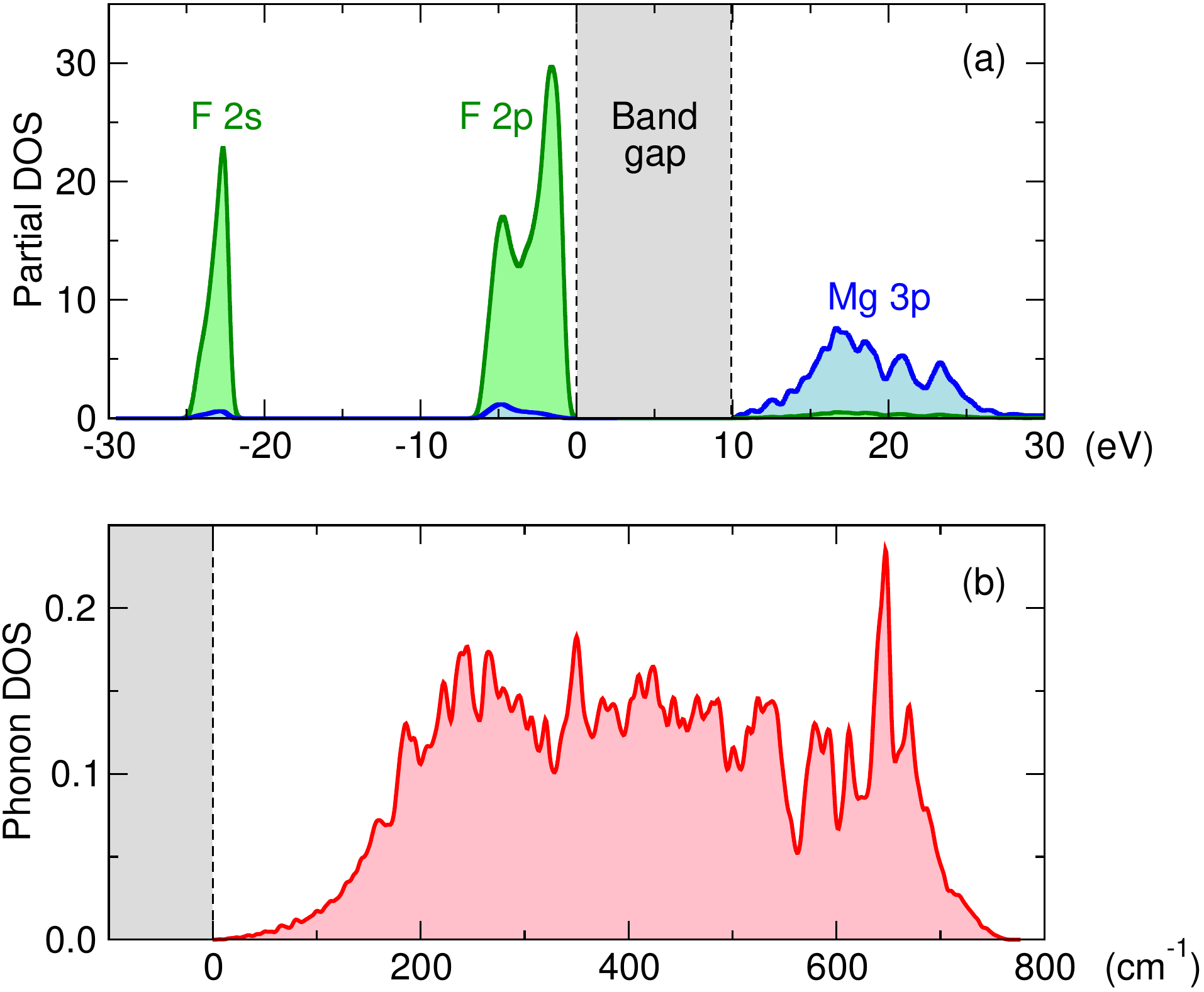}}\quad
  \subfigure{\includegraphics[scale=0.38,clip]{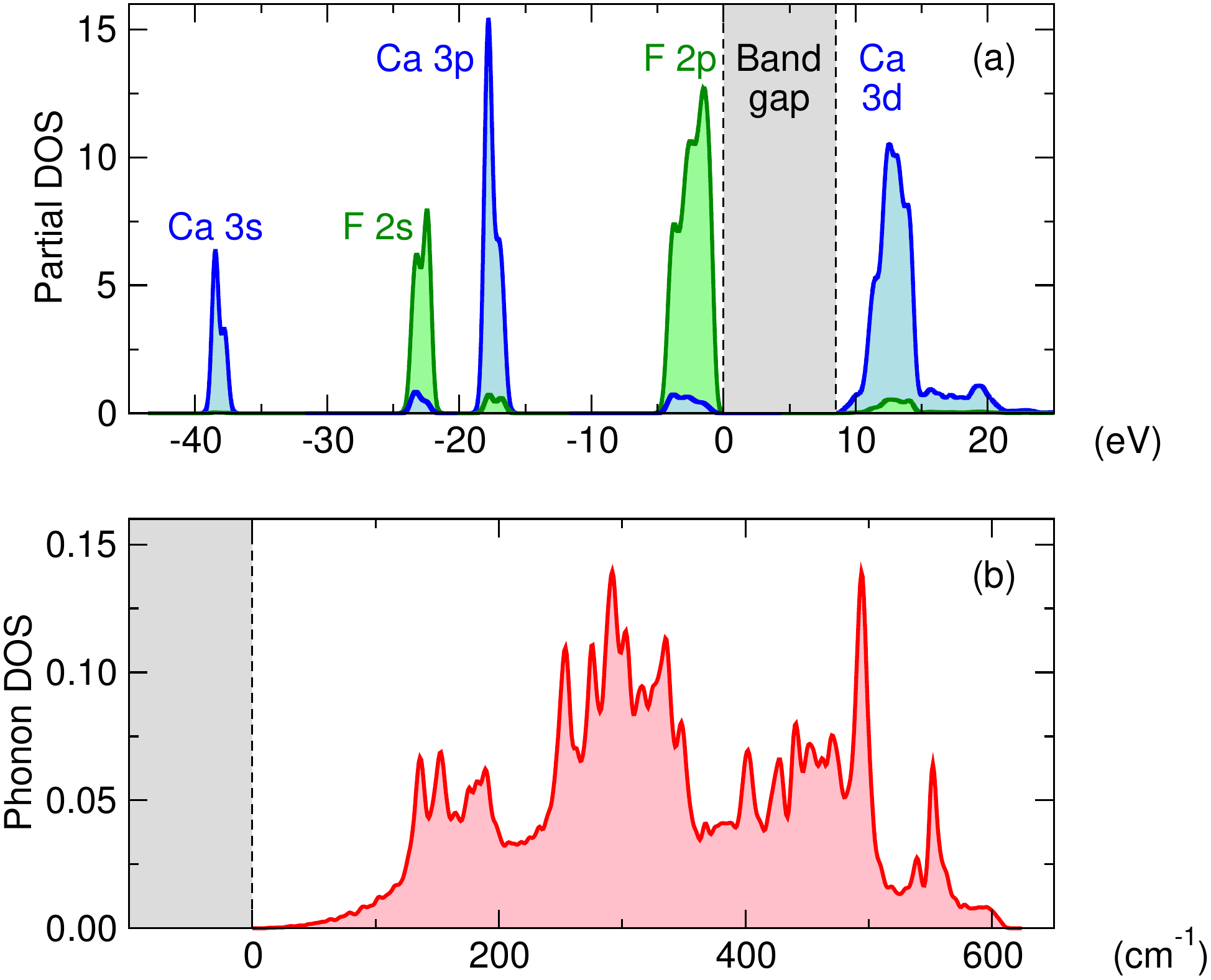}}
  \caption{\label{fig:Pbca_and_P62m} Electronic (a) and phonon (b) density of states for the O-I$-$MgF$_2$ phase at 42 GPa (left panel), and for the $P\overline{6}2m$-CaF$_2$ phase at 30 GPa (right panel). A 4$\times$3$\times$2 (576 atom) and 3$\times$3$\times$6 (486 atom) supercell is used for the O-I and $P\overline{6}2m$ phonon calculations, respectively.}
\end{figure*}

In this section we show the electronic and phonon
density$-$of$-$states (DOS) for some of the newly predicted structures
discussed in the main paper. Specifically, these are the
moganite$-$BeF$_2$ and CaCl$_2$$-$BeF$_2$
structures, stable in the pressure ranges 11.6$-$30.1
GPa and 30.1$-$57.5
GPa respectively (Fig.~\ref{fig:moganite_and_cacl2}), and the
O-I$-$MgF$_2$
and $P\overline{6}2m$-CaF$_2$
structures (Fig.~\ref{fig:Pbca_and_P62m}). The O-I$-$MgF$_2$
structure is predicted to be stable in the pressure range
39.6$-$44.1
GPa, while the $P\overline{6}2m$-CaF$_2$
phase is calculated to become stable at high temperature.

\vspace{0.2cm}

The electronic DOS ((a) in the figures below) is calculated with the
\textsc{Optados} code using the same DFT parameters as our bandgap
calculations (see Sec.~\ref{sec:esc} above) and a Gaussian smearing of
0.25 eV. The valence band maximum (VBM) is set to 0 eV. The majority
contributor (species and orbital) to each feature in the DOS is
labelled in Figs.~\ref{fig:moganite_and_cacl2} and
\ref{fig:Pbca_and_P62m} below. In all cases, we find that the valence
bands consist of fluorine 2$p$ orbitals, while the conduction bands
come from unoccupied $p$ or $d$ orbitals in beryllium, magnesium or
calcium.

\vspace{0.2cm}

The dynamic stability of the predicted structures is indicated by the
phonon DOS ((b) in the figures below) for these structures having no
protrusion into negative (imaginary) frequencies (shown as grey
regions).

\section{Known AB$_2$ structures used}
A list of $AB_2$-type structures that were considered in addition to
our AIRSS searches is given below. The atoms are substituted
appropriately, for example Si$\rightarrow$Be, O$\rightarrow$F, and the
structure relaxed. Structures are taken from the Inorganic Crystal
Structure Database (ICSD) \cite{ICSD-II} or from literature. We give the
5-digit ICSD reference number for the former.

\vspace{0.2cm}

In the table below, $Z$ is the number of formula units in the
\textit{primitive} unit cell, and so $3Z$ is the total number of atoms
in the primitive cell. Our AIRSS searches were restricted to
$Z\leq 8$.

\vspace{0.2cm}

\begin{center}
\begin{tabular}{| l | l | l | l | l | l | l |} \hline
Compound & SG (HM) & SG (\#) & Lattice & $Z$ & Name(s) & Reference \\ \hline
HfO$_2$  & $P2_1/c$  & 14 & monoclinic   & 4   & Baddeleyite  & 60903   \\
SiO$_2$  & $P2_1/c$  &  14 & monoclinic  & 32 & Coesite-II      & \cite{Cernok-2014-II} \\
SiO$_2$  & $C2/c$    & 15  & monoclinic  & 8  & Coesite-I       & \cite{Cernok-2014-II} \\
SiO$_2$  & $C2/c$    & 15  & monoclinic  & 6   & Moganite      & 67669 \\
BeF$_2$  & $C2/c$    & 15  & monoclinic & 6   & $-$           & \cite{Rakitin-2015-II} \\ \hline
SiO$_2$  & $Pnnm$    & 58  & orthorhombic  & 2   & CaCl$_2$-type  &  26158 \\
SiO$_2$  & $Pbcn$    & 60 & orthorhombic  & 4   & $\alpha$-PbO$_2$ $^{(a)}$ & \cite{ElGoresy-2008} \\
TiO$_2$  & $Pbca$    & 61  & orthorhombic  & 8   & Orthorhombic-I, OI & \cite{Dubrovinskaia-2001-II} \\ 
ZrO$_2$  & $Pnma$    & 62  & orthorhombic  & 4   & Cotunnite, Ortho-II (PbCl$_2$) & \cite{Lowther-1999} \\ \hline
SiO$_2$  & $P4_12_12$   & 92 & tetragonal  &  4  & $\alpha$-cristobalite & 75300 \\
SiO$_2$  & $P4_2/mnm$   & 136  & tetragonal  & 2   & Rutile, $\beta$-PbO$_2$, Plattnerite  & 9160 \\ \hline
SiO$_2$  & $P3_121$     & 152  & trigonal & 3   & $\alpha$-quartz & 27745 \\
SiO$_2$  & $P6_222$     & 180  & hexagonal & 3   & $\beta$-quartz  & 31088 \\ \hline
SiO$_2$  & $Pm\overline{3}m$ & 221 & cubic & 46 & Melanophlogite & 159046$^{(b)}$ \\ 
SiO$_2$  & $Fd\overline{3}m$ & 227 & cubic  & 2  & $\beta$-cristobalite & 77458 \\ \hline
\multicolumn{7}{l}{$^{(a)}$ Has a number of aliases: Seifertite (SiO$_2$), Scrutinyite ($\alpha$-PbO$_2$), Columbite (TiO$_2$)} \\
\multicolumn{7}{l}{$^{(b)}$ Interstitial carbon atoms (representing methane) were removed from this structure before relaxing} \\
\end{tabular}
\end{center}

\vspace{0.2cm}

For the orthorhombic space groups, a number of different settings are possible with different HM symbols. We list some of the orthorhombic ones below \cite{bilbao}:

\vspace{0.2cm}

\begin{center}
\begin{tabular}{| l | l | l |} \hline
SG (\#) & Standard setting & Alternative settings \\ \hline
58   & $Pnnm$  & $Pmnn$, $Pnmn$ \\
60   & $Pbcn$  & $Pcan$, $Pnca$, $Pnab$, $Pbna$, $Pcnb$ \\
61   & $Pbca$  & $Pcab$ \\
62   & $Pnma$  & $Pmnb$, $Pbnm$, $Pcmn$, $Pmcn$, $Pnam$ \\ \hline
\end{tabular}
\end{center}

\newpage

\section{Quasiharmonic phase diagram for CaF$_2$}

\subsection{Computational details}
To construct the phase diagram shown in Fig.~8 of the main paper, we
perform finite-displacement phonon calculations at several different
volumes for the $Fm\overline{3}m$, $P\overline{6}2m$ and $Pnma$
CaF$_2$ structures. We use 5$\times$5$\times$5 (375 atom),
3$\times$3$\times$6 (486 atom), and 5$\times$3$\times$3 (540 atom)
supercells for the $Fm\overline{3}m$, $P\overline{6}2m$ and $Pnma$
structures, respectively. Volumes corresponding to the following
static-lattice pressures are used for these calculations:

\vspace{0.2cm}

$Fm\overline{3}m$: $-$6, $-$4, $-$2, 0, 4, 8, 12, 16, 20 GPa

$P\overline{6}2m$: 10, 20, 30, 40, 50, 60, 70 GPa

$Pnma$: 0, 2, 6, 10, 20, 30, 40, 50, 60, 70 GPa

\vspace{0.2cm}

This gives us the phonon free energies $F(V,T)$ at each volume for these structures.

\begin{figure*}[htp]
  \centering
  \subfigure{\includegraphics[scale=0.28,clip]{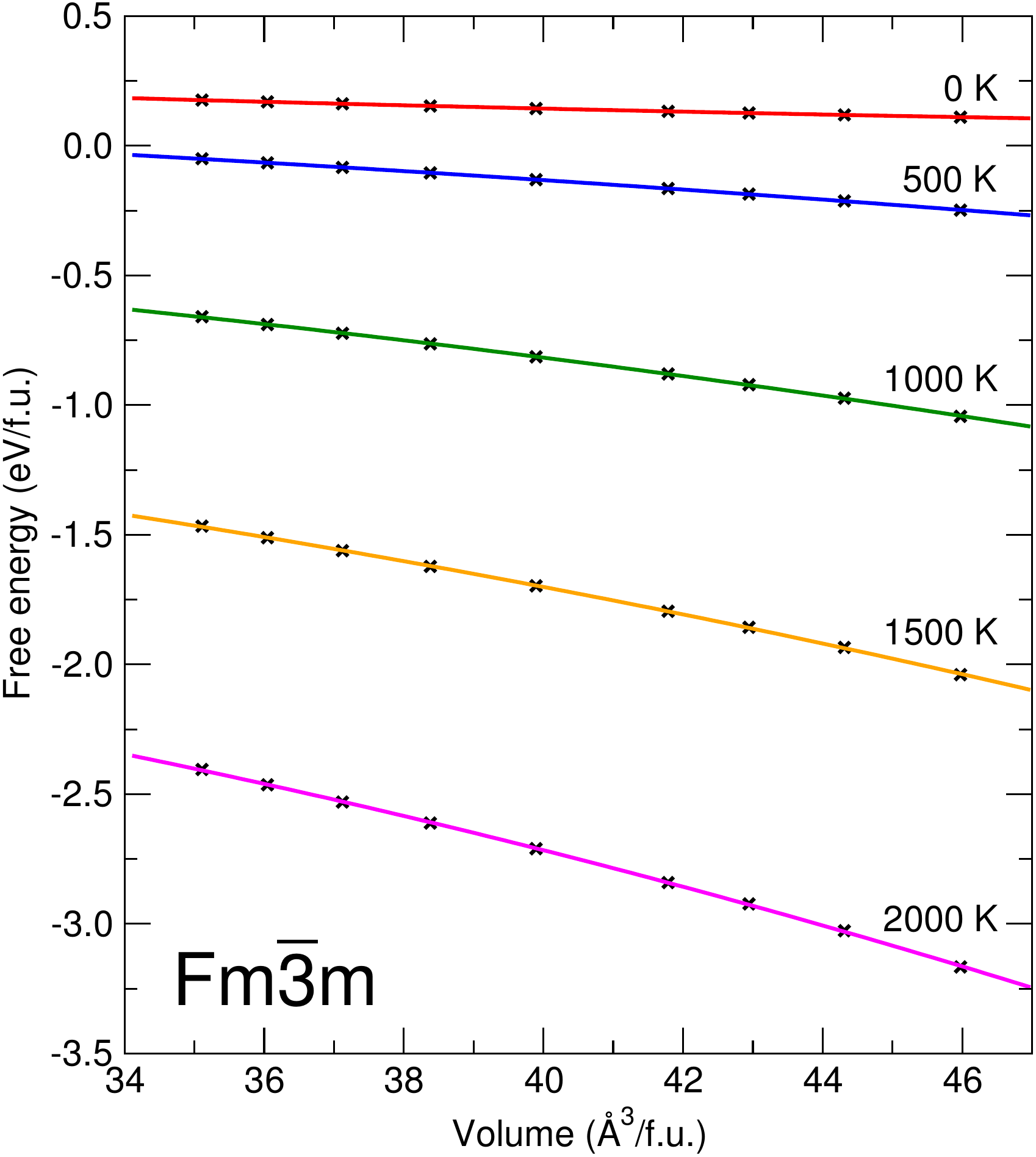}}\quad
  \subfigure{\includegraphics[scale=0.28,clip]{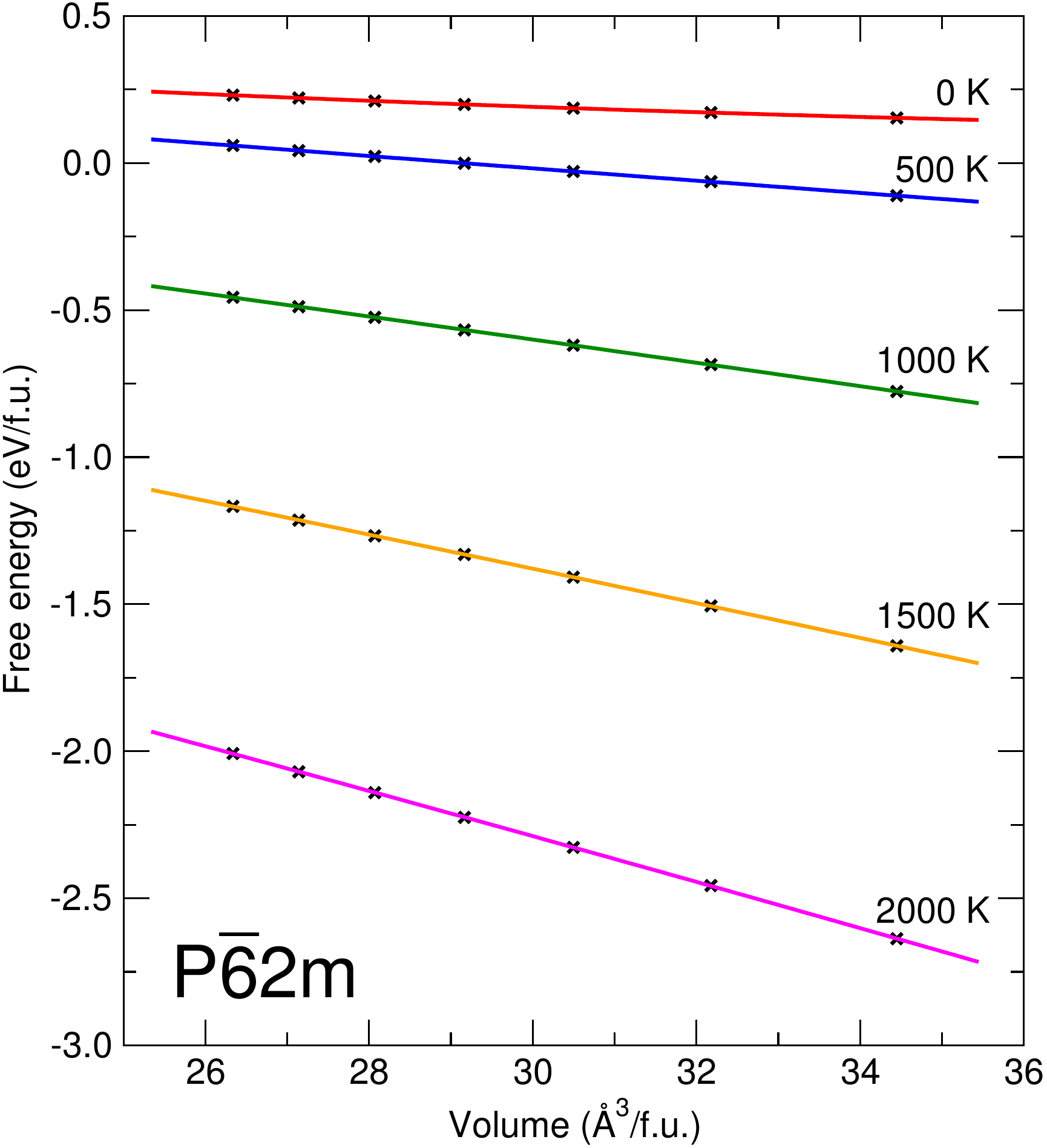}}\quad
  \subfigure{\includegraphics[scale=0.28,clip]{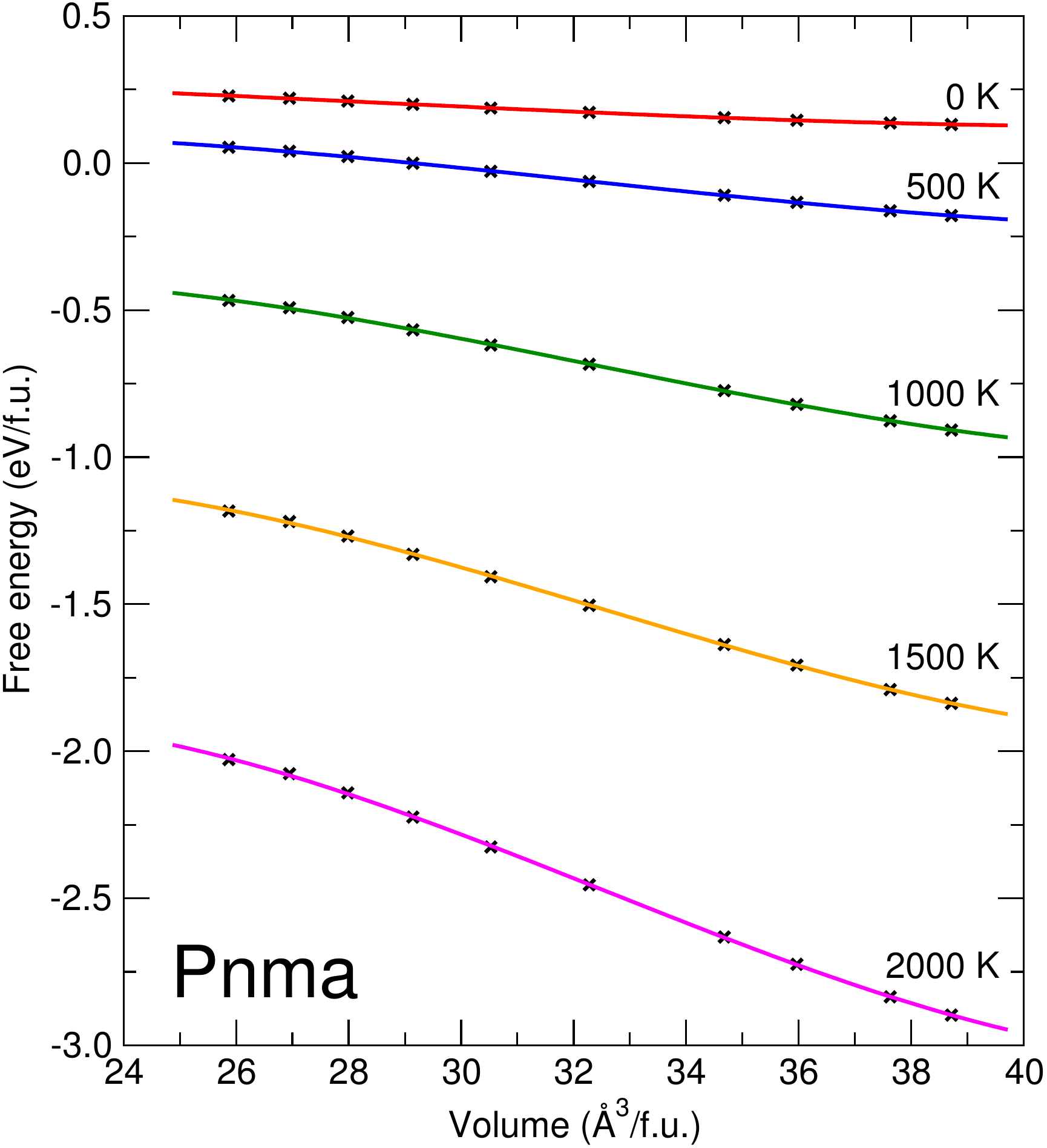}}
  \caption{\label{fig:CaF2_free_energies} Phonon free energies $F(V)$ for the $Fm\overline{3}m$ (left), $P\overline{6}2m$ (middle) and $Pnma$ (right) structures of CaF$_2$, as a function of volume and of temperature. The `+' marks indicate values from supercell calculations, while the solid curves show the resulting polynomial fits to $F(V)$.}
\end{figure*}

\begin{figure*}[htp]
  \centering
  \subfigure{\includegraphics[scale=0.28,clip]{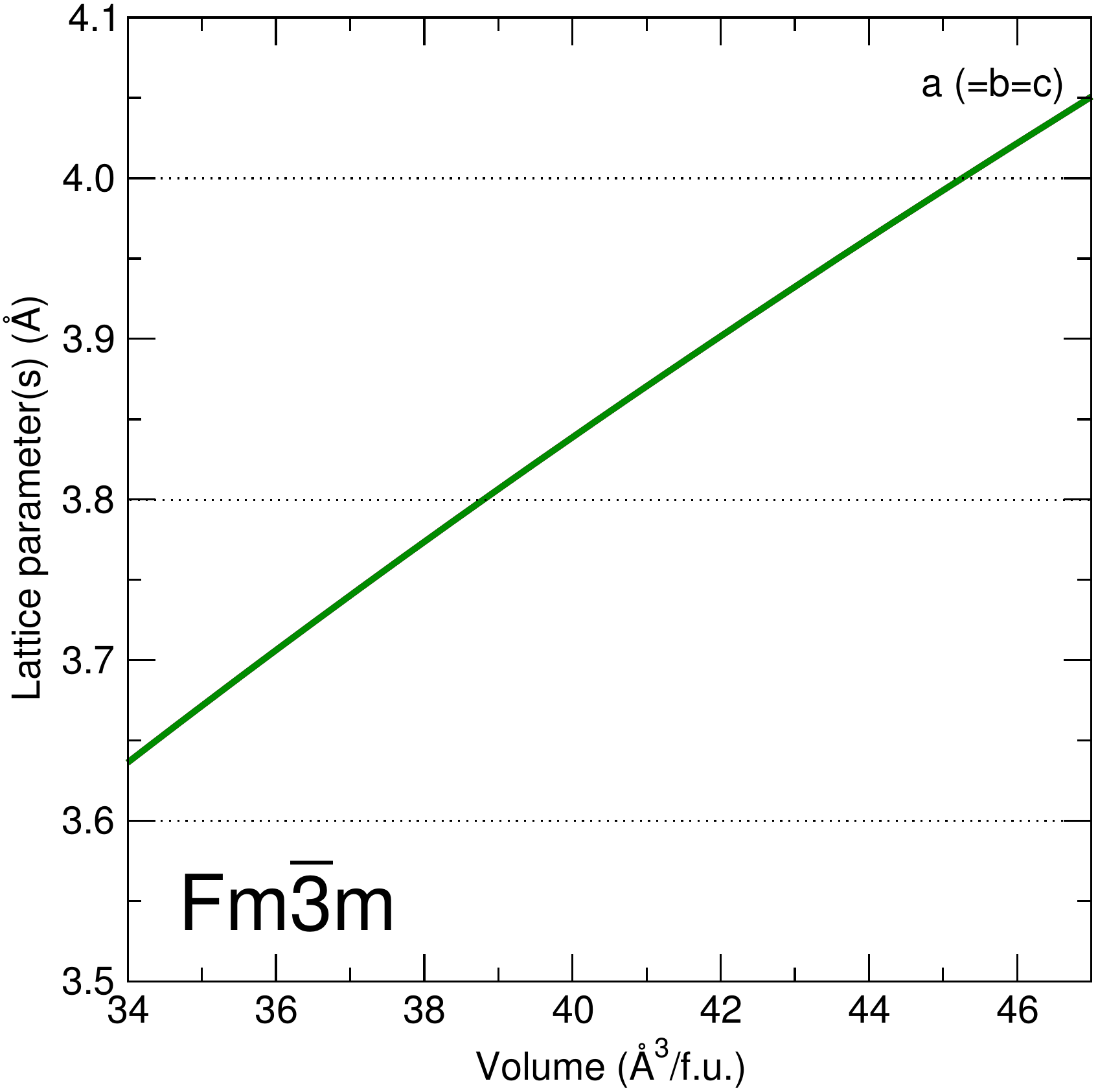}}\quad
  \subfigure{\includegraphics[scale=0.28,clip]{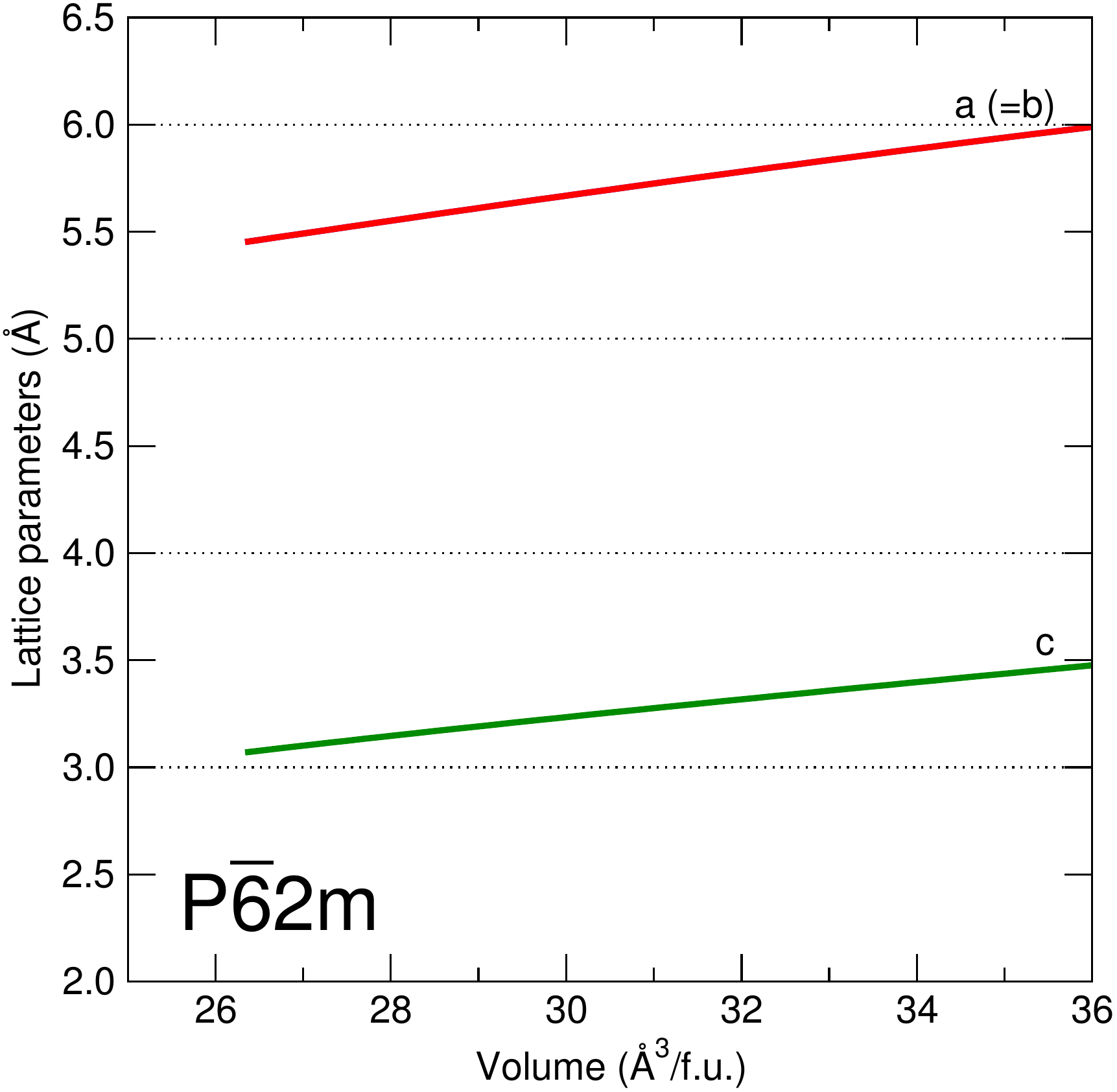}}\quad
  \subfigure{\includegraphics[scale=0.28,clip]{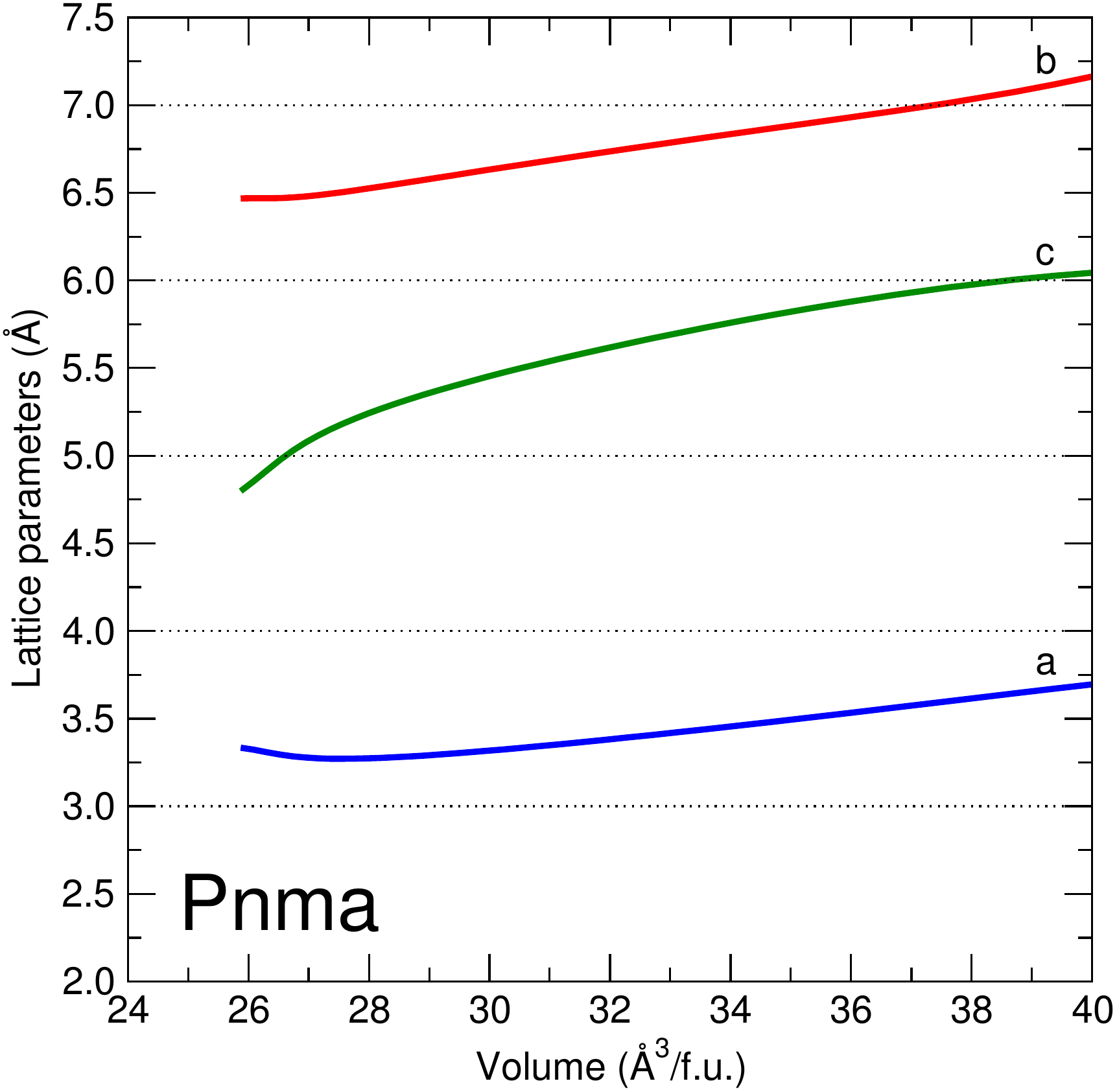}}
  \caption{\label{fig:CaF2_latticeparams} Lattice parameters of the $Fm\overline{3}m$ (left), $P\overline{6}2m$ (middle) and $Pnma$ (right) structures of CaF$_2$ as a function of volume, calculated at the static lattice level. The lattice parameter for $Fm\overline{3}m$ is given for the primitive cell ($Z$=1, $a$=$b$=$c$, $\alpha$=$\beta$=$\gamma$=$60^{\circ}$); the lattice parameter for the corresponding conventional cell ($Z$=4) is $\sqrt{2}$ times larger. Our simulation cell for the $Pnma$ structure has $a<c<b$, and is actually set in the alternative $Pmcn$ setting for that space group (\#62).}
\end{figure*}

We fit the resulting energies with a polynomial in $V$. For the
$Fm\overline{3}m$ and $P\overline{6}2m$ structures, the calculated
$F(V,T)$ values vary reasonably slowly with volume and are well
reproduced by a quadratic in $V$. For the $Pnma$ structure, we find
that the calculated $F(V)$ values tend to initially increase rapidly
with decreasing volume, but then increase less rapidly at lower
volumes (see Fig.~\ref{fig:CaF2_free_energies}). A cubic polynomial in
$V$ is therefore used to capture this trend. Some insight into this
behaviour can be found by looking at the dependence of the lattice
parameters of these structures on $V$; for $Fm\overline{3}m$ and
$P\overline{6}2m$, the lattice parameters steadily decrease on
compression. The $Pnma$ structure, however, shows slightly unusual
behaviour on compression in that lattice parameter $a$ initially
decreases, but then starts increasing for static-lattice pressures
above 56 GPa, while lattice parameter $c$ decreases more rapidly at
that pressure as well (Fig.~\ref{fig:CaF2_latticeparams}).

\vspace{0.2cm}

Static-lattice calculations provide relaxed geometries for our
calculations, and give us sets of enthalpies $H$, volumes $V$ and
static-lattice pressures (or electron pressures) $P_{st}$. Within the
QHA, at a given volume $V$, the corresponding Gibbs free energy is the
sum of the electronic enthalpy and phonon free energy:
$G(V,T) = H(V) + F(V,T)$. The pressure $P$ at that volume is
$P = P_{st} + P_{ph}$, where
$P_{ph} \equiv -\partial F(V,T)/\partial V$ is the phonon pressure.

\subsection{Validity of the QHA}
\begin{figure*}[htp]
  \centering
  \subfigure{\includegraphics[scale=0.28,clip]{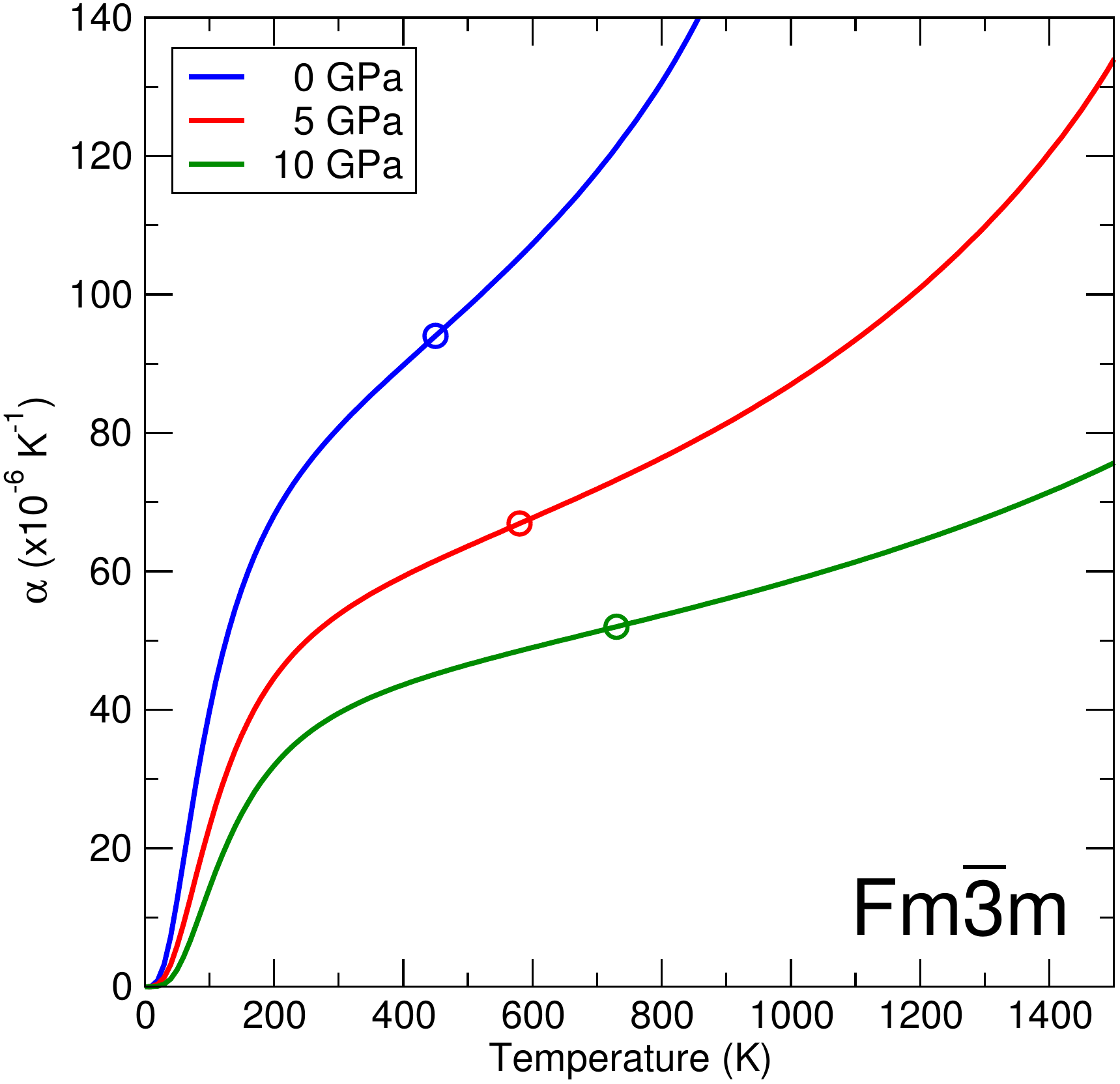}}\quad
  \subfigure{\includegraphics[scale=0.28,clip]{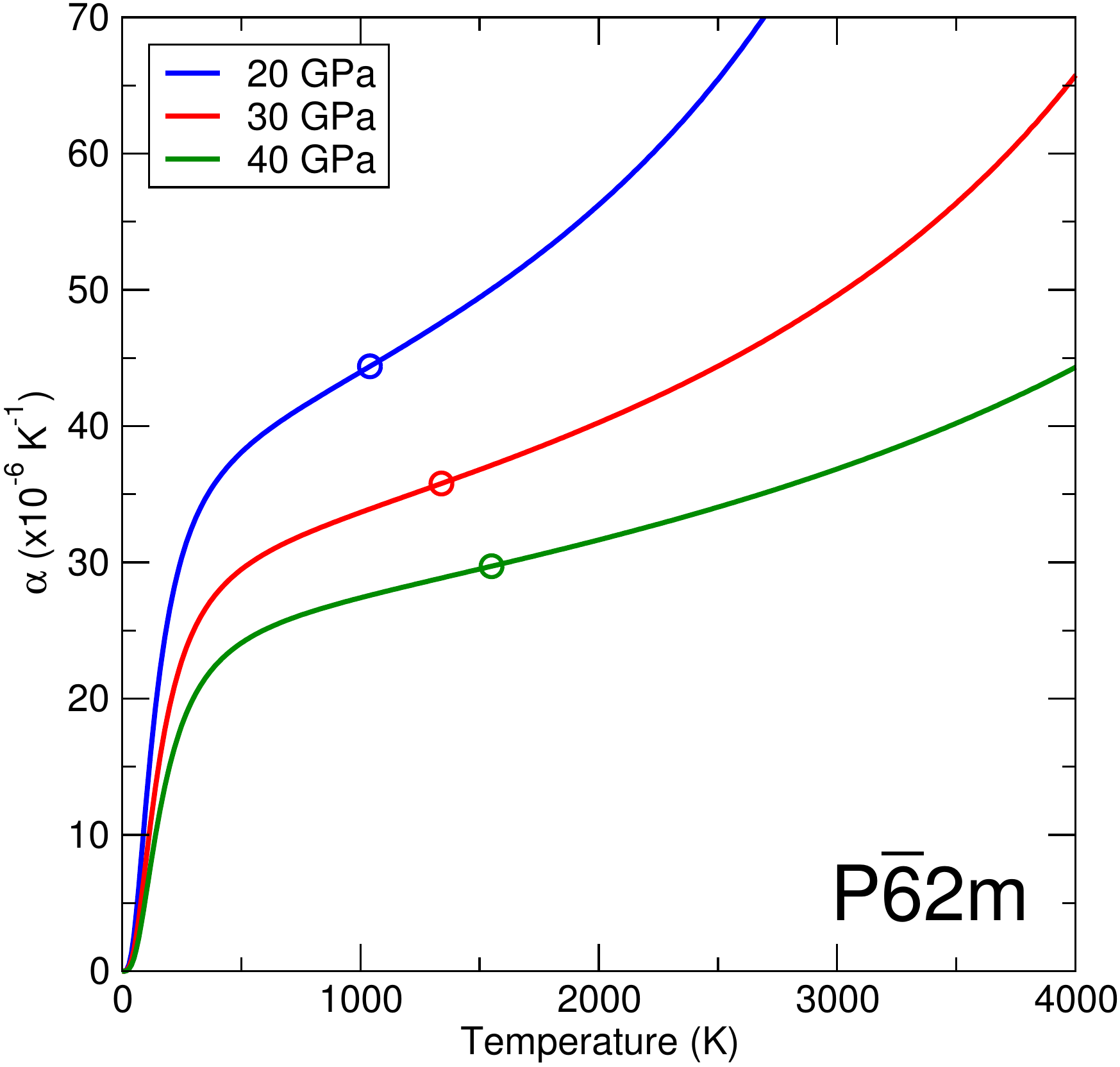}}\quad
  \subfigure{\includegraphics[scale=0.28,clip]{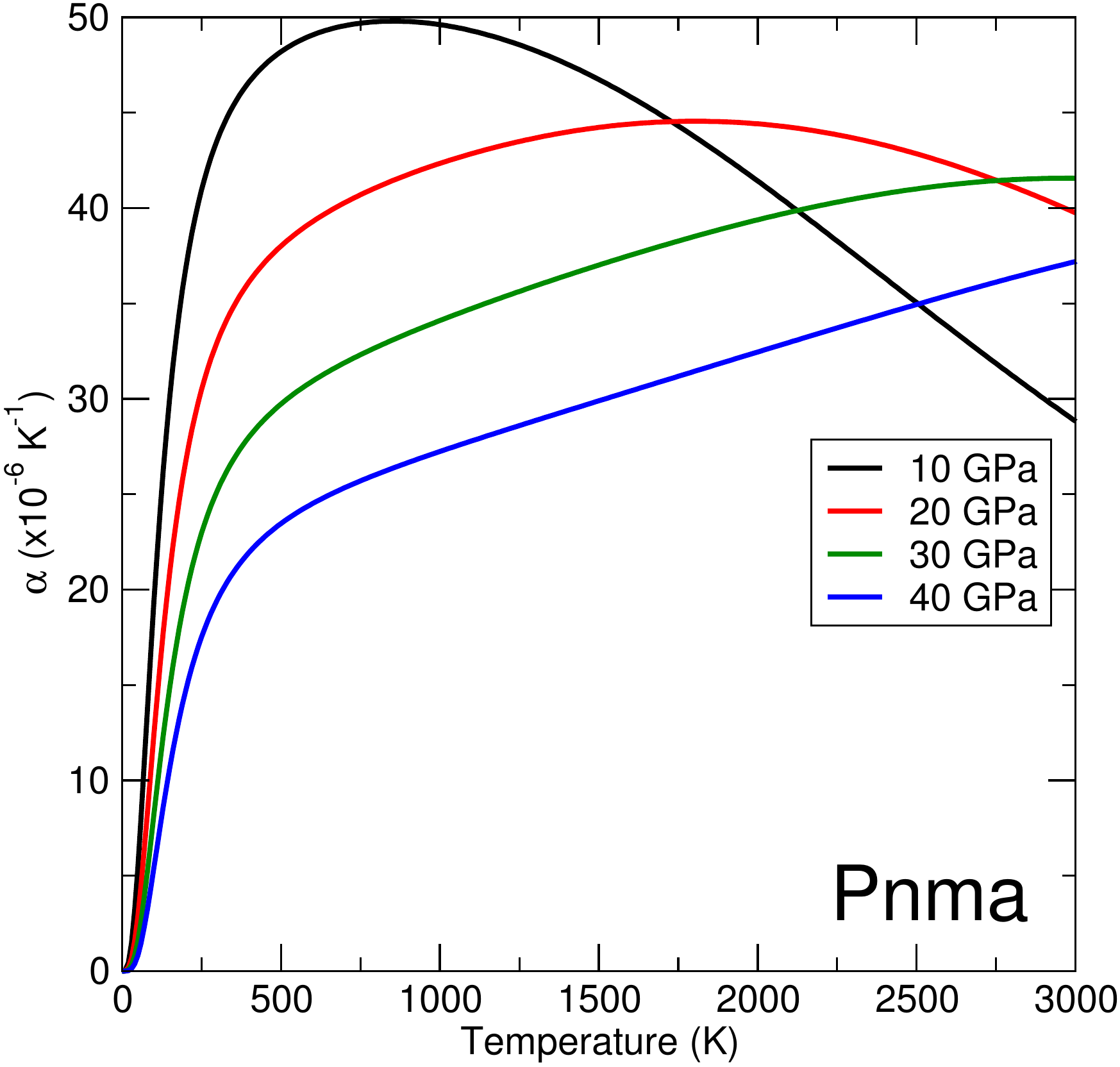}}
  \caption{\label{fig:CaF2_thermale} Calculated volume thermal expansion coefficients $\alpha (P,T)$ at a variety of pressures as a function of temperature for the $Fm\overline{3}m$ (left), $P\overline{6}2m$ (middle) and $Pnma$ (right) CaF$_2$ structures. The open circles in the left and middle panels show inflection points, where $d^2\alpha/dT^2$ changes sign.}
\end{figure*}

Here we calculate the volume coefficient of thermal expansion:
\begin{eqnarray*}
\alpha(P,T) = \frac{1}{V}\left(\frac{\partial V}{\partial T}\right)_P
\end{eqnarray*}

at a variety of pressures for the $Fm\overline{3}m$,
$P\overline{6}2m$ and $Pnma$ calcium fluoride structures. The results
are shown in Fig.~\ref{fig:CaF2_thermale}.

\vspace{0.2cm}

The calculated thermal expansion coefficient can be used to assess the validity of the QHA \cite{Karki-2001-II,Wentzcovitch-2004-II}. We can consider the QHA valid at temperatures lower than the higher temperature inflection point in $\alpha(P,T)$, i.e. where
\begin{eqnarray*}
\left[\frac{\partial^2 \alpha(P,T)}{\partial T^2}\right]_P 
\end{eqnarray*}

changes sign. Beyond this point, we generally find that
$\left[\frac{\partial^2 \alpha(P,T)}{\partial T^2}\right]_P \geq 0$
and the QHA is expected to be less applicable with further increases
in temperature, as anharmonic effects become more important.

\vspace{0.2cm}

For the $Fm\overline{3}m$ and $P\overline{6}2m$ structures,
Fig.~\ref{fig:CaF2_thermale} shows this inflection point in
$\alpha(P,T)$ using open circles at each pressure. Fitting the
resulting $(P,T)$ coordinates of these inflection points gives us the
validity rule $T \mbox{ (}K\mbox{)}\leq 28P\mbox{ (GPa)} + 453$ given
in the main paper. We note that locating the inflection point is
subject to some numerical uncertainty, because $\alpha(P,T)$ changes
quite slowly in the vicinity of this point
(Fig.~\ref{fig:CaF2_thermale}). An uncertainty of about 100 K is
therefore quoted in the main paper.

\vspace{0.2cm}

This inflection point is harder to locate in the $\alpha(P,T)$ curves
for the $Pnma$ structure. The thermal expansion coefficients are
slightly unusual in that they initially increase, but then decline at
higher temperatures. This is partly due to the fact that the phonon
pressure in $Pnma$ declines at higher volumes, as can be seen in the
slope of the $F(V,T)$ curves for $Pnma$ in the right$-$hand panel of
Fig.~\ref{fig:CaF2_free_energies}.

\section{Unstable modes in $Fm\overline{3}m$ and $P\overline{6}2m$ CaF$_2$}

\begin{figure*}[htp]
  \centering
  \subfigure{\includegraphics[scale=0.4,clip]{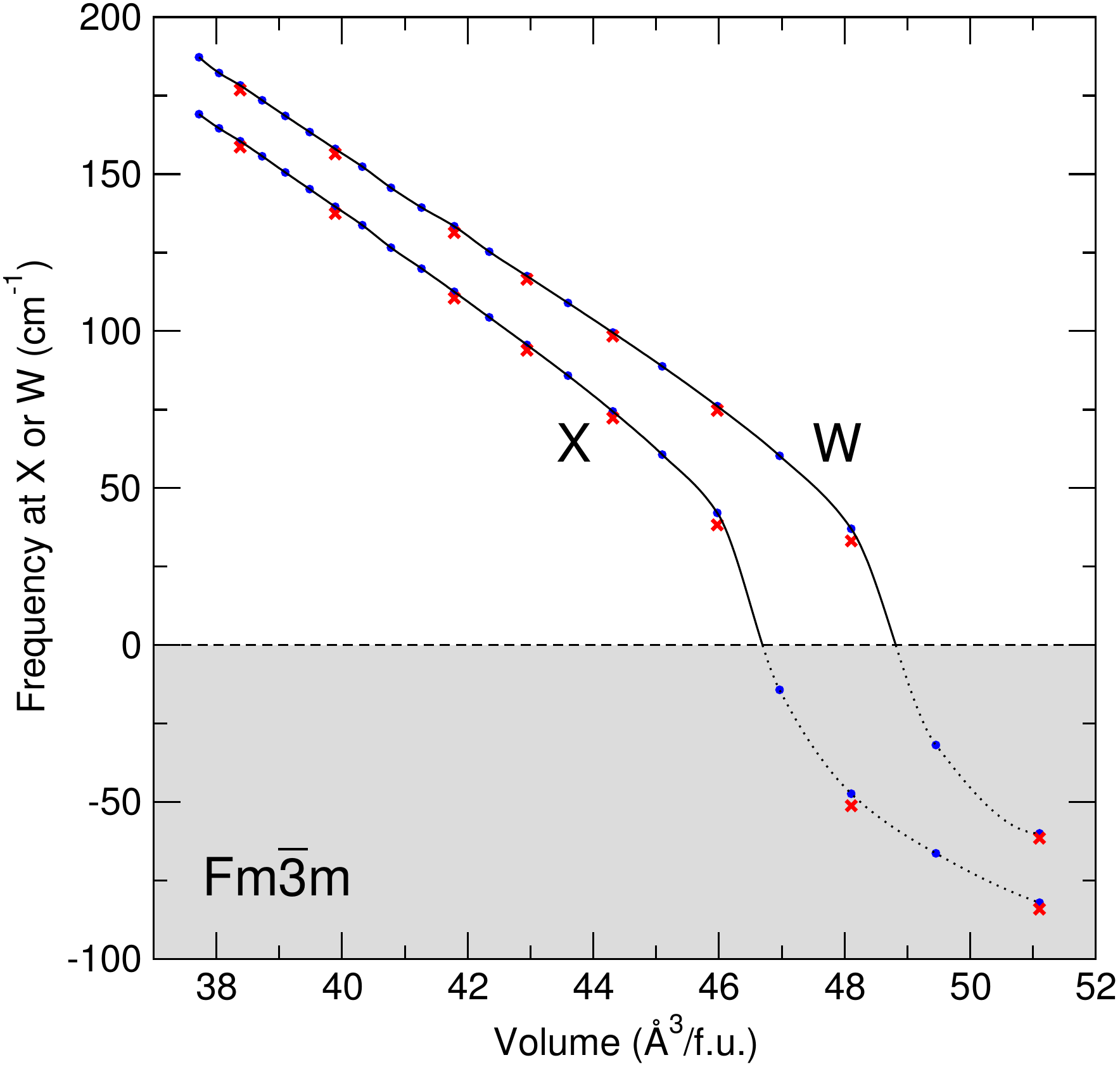}}\quad
  \subfigure{\includegraphics[scale=0.4,clip]{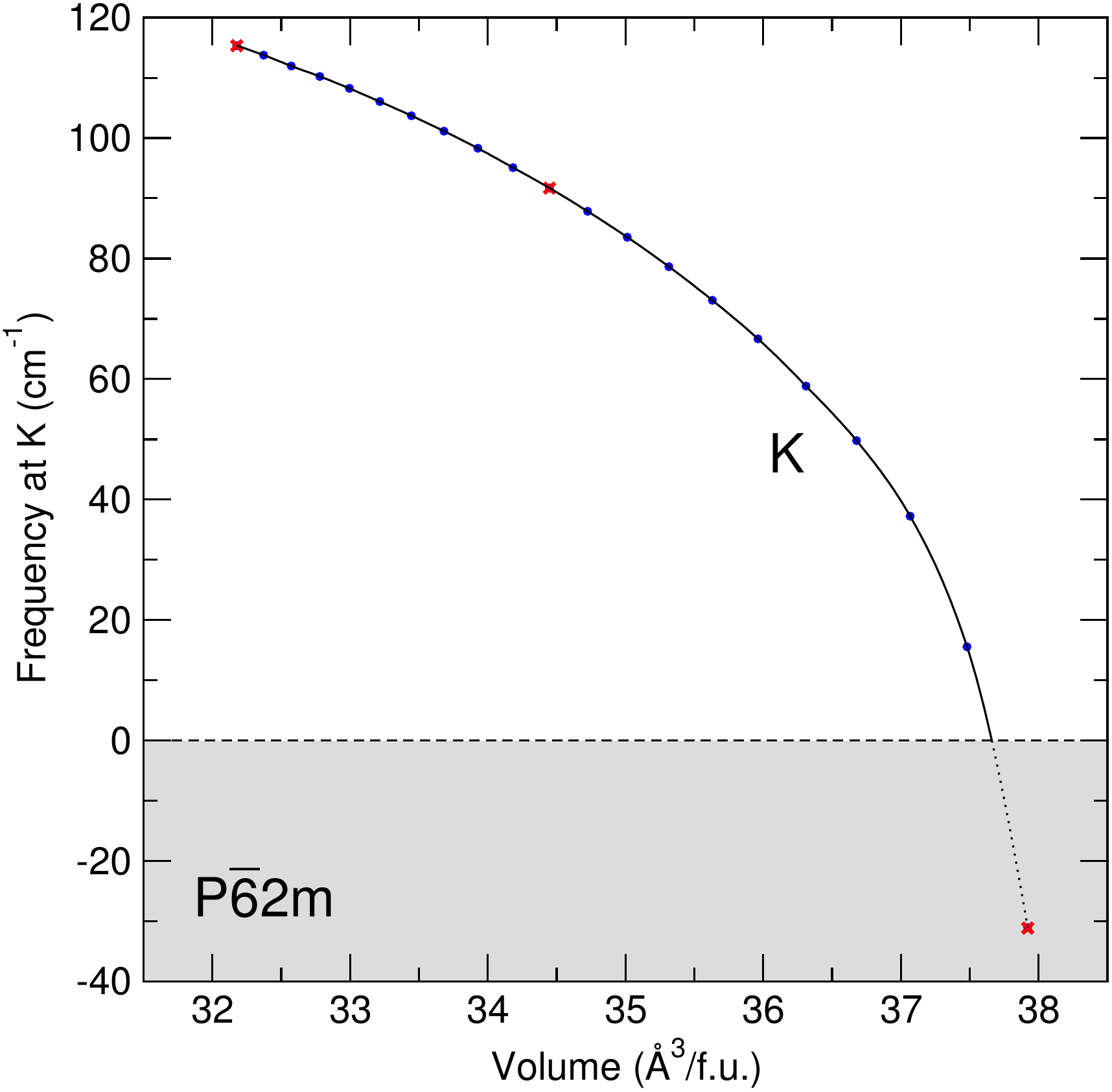}}
  \caption{\label{fig:CaF2_Fm3m_P62m_unstablelatticeparams} (Left) Lowest mode frequency at the Brillouin zone $X$ (left$-$hand curve) and $W$ (right$-$hand curve) points for $Fm\overline{3}m$-CaF$_2$, as a function of volume. (Right) Mode frequencies at the Brillouin zone $K$ point for $P\overline{6}2m$-CaF$_2$ as a function of volume. Imaginary frequencies are shown as negative values.}
\end{figure*}

In Figs.~9 and 10 of the main paper, the $Fm\overline{3}m$ and
$P\overline{6}2m$ structures develop unstable phonon frequencies with
decreasing pressure / increasing volume. For the $Fm\overline{3}m$
structure, imaginary (`negative') phonon frequencies first develop at
the Brillouin zone $X$ point, and then at larger volumes at the $W$
point. Fig.~\ref{fig:CaF2_Fm3m_P62m_unstablelatticeparams} shows the
smallest mode frequency at $X$ and $W$ as a function of volume for
this structure. The solid blue points in
Fig.~\ref{fig:CaF2_Fm3m_P62m_unstablelatticeparams} were obtained
using a 2$\times$4$\times$4 supercell (96 atoms) for
$Fm\overline{3}m$, as the $X$ and $W$ points are both commensurate
with a supercell of this size. For comparison, red crosses in
Fig.~\ref{fig:CaF2_Fm3m_P62m_unstablelatticeparams} show the results
obtained using a 5$\times$5$\times$5 supercell (375 atoms), the size
used in our phase diagram calculations. The $X$ and $W$ points are not
commensurate with this supercell, so the frequencies shown with red
crosses have been obtained by interpolation. These results differ from
the exactly commensurate 2$\times$4$\times$4 supercell results by $<$4
cm$^{-1}$. The thin solid and dotted black lines connecting the blue
points are shape$-$preserving interpolants fitted to the data and are
intended to guide the eye.

\vspace{0.2cm}

For the $P\overline{6}2m$ structure, imaginary phonons develop at the
Brillouin zone $K$ point, and the right$-$hand panel of
Fig.~\ref{fig:CaF2_Fm3m_P62m_unstablelatticeparams} also shows the
lowest mode frequency at this point as a function of volume. The solid
blue points are obtained using a 3$\times$3$\times$1 supercell (81
atoms) commensurate with $K$, while the red crosses show results using
a 3$\times$3$\times$6 supercell (486 atoms), as used in phase diagram
calculations. As with the $Fm\overline{3}m$ structure, the red crosses
are interpolations; however, we note that $K$ is also commensurate
with this larger supercell, and the interpolated results are almost
identical to the exactly commensurate ones in this case. The thin
solid and dotted black lines are again guides to the eye.

\vspace{0.2cm}

The volume at which these structures first develop imaginary phonon
frequencies directly affects the phase diagram of Fig. 8 of the main
paper. In particular, this volume determines the location of the
dashed lines between the red/light$-$red and blue/light$-$blue regions
of the phase diagram. We have taken this volume to be simply halfway
between the first blue point in
Fig.~\ref{fig:CaF2_Fm3m_P62m_unstablelatticeparams} with a frequency
$>$0 cm$^{-1}$, and the first `negative-frequency' blue point. For
$Fm\overline{3}m$, this corresponds to \mbox{$V= 46.47$ \AA$^3$/f.u.},
and is at a static$-$lattice pressure between $-7$ and $-6$ GPa. For
$P\overline{6}2m$, we have \mbox{$V= 37.70$ \AA$^3$/f.u.}, occurring
between 0 and 1 GPa.

\end{document}